\documentclass[10pt,aps,prd,amsmath,amssymb,nofootinbib,superscriptaddress,preprintnumbers,hidelinks,eprint,reprint,
nofootinbib,nobibnotes,
floatfix
]{revtex4-2}
\usepackage{ragged2e}
\usepackage{etoolbox}
\usepackage{orcidlink}
\usepackage{tabularray}
\usepackage{graphicx}% Include figure files
\usepackage{dcolumn}% Align table columns on decimal point
\usepackage{bm}% bold math
\usepackage{cancel}
\usepackage{array}
\usepackage{array,url,kantlipsum}

\usepackage{amssymb}

\usepackage{mathabx}
\usepackage{mathrsfs}  
\usepackage{tensor}
\usepackage{comment}
\usepackage[normalem]{ulem}
\usepackage{cancel}
\usepackage{booktabs}
\usepackage[caption=false]{subfig}

\usepackage{graphicx}
\usepackage{dcolumn}
\usepackage{xcolor,colortbl}
\usepackage{multirow}
\usepackage{comment}
\usepackage{enumitem}

\usepackage{multirow}

\usepackage{hyperref}
\usepackage{stackengine,scalerel}
\hypersetup{colorlinks, linkcolor={red},citecolor={blue},urlcolor={blue}}  

\usepackage{physics,mathtools,tensor}
\usepackage{caption,subcaption} 
\usepackage{soul}
\usepackage{verbatim}

\newcommand{\fHn}[1]{f_H^{(#1)}}
\newcommand{\hHn}[1]{h_H^{(#1)}}
\newcommand{\rps}{r_\text{ps}}

\newcommand{\bsh}{b_\text{sh}}
\newcommand{\hr}{r_H}

\newcommand{\newom}{\Omega}
\newcommand{\newgam}{\Gamma}

\def\bcond{\left\vert_{\begin{array}{l}\scriptstyle r=\hr \end{array}}\right.\hskip-0.1cm}

\renewenvironment{widetext@grid}{%
  \par\ignorespaces
  \setbox\widetext@top\vbox{%
   \vskip15\p@
   \hb@xt@\hsize{%
    \leaders\hrule\hfil
    \vrule\@height6\p@
   }%
   \vskip6\p@
  }%
  \setbox\widetext@bot\hb@xt@\hsize{%
    \vrule\@depth6\p@
    \leaders\hrule\hfil
  }%
  \onecolumngrid
  \let\set@footnotewidth\set@footnotewidth@ii
}{%
  \par
  \twocolumngrid\global\@ignoretrue
  \@endpetrue
}%
\makeatother
\begin{document}

\title{\bf 
Toward a unified view of agnostic parametrizations for deformed black holes}

\author{Manuel Del Piano {\Large \orcidlink{0000-0003-4515-8787}}\,}
\email[]{manuel.delpiano-ssm@unina.it}
\affiliation{Scuola Superiore Meridionale, Largo S. Marcellino, 10, 80138 Napoli, Italy.}
\affiliation{INFN sezione di Napoli, via Cintia, 80126 Napoli, Italy.}
\affiliation{Quantum  Theory Center ($\hbar$QTC) \& D-IAS, Southern Denmark University, Campusvej 55, 5230 Odense M, Denmark.}

\author{Ciro De Simone {\Large \orcidlink{0009-0004-0610-1686}}\,}
\email[]{ciro.desimone@unina.it}
\affiliation{Department of Physics E. Pancini, Università di Napoli Federico II, via Cintia, 80126 Napoli, Italy.}
\affiliation{INFN sezione di Napoli, via Cintia, 80126 Napoli, Italy.}

\author{Mattia Damia Paciarini {\Large \orcidlink{0009-0000-1044-341X}}\,}
\email[]{damiapaciarinim@qtc.sdu.dk}
\affiliation{Quantum  Theory Center ($\hbar$QTC) \& D-IAS, Southern Denmark University, Campusvej 55, 5230 Odense M, Denmark.}

\author{Vittorio De Falco {\Large \orcidlink{0000-0002-4728-1650}}\,}
\email[]{deltafi.mat@live.it}
\affiliation{Ministero dell'Istruzione e del Merito (M.I.M., ex M.I.U.R.)}

\author{Mikołaj Myszkowski {\Large \orcidlink{0000-0002-5207-4509}}\,}
\email[]{mikolaj@qtc.sdu.dk}
\affiliation{Quantum  Theory Center ($\hbar$QTC) \& D-IAS, Southern Denmark University, Campusvej 55, 5230 Odense M, Denmark.}

\author{Francesco Sannino {\Large\orcidlink{0000-0003-2361-5326}}\,}
\email[]{sannino@qtc.sdu.dk}
\affiliation{Quantum Theory Center ($\hbar$QTC) \& D-IAS, Southern Denmark University, Campusvej 55, 5230 Odense M, Denmark.}
\affiliation{Scuola Superiore Meridionale, Largo S. Marcellino, 10, 80138 Napoli, Italy.}
\affiliation{INFN sezione di Napoli, via Cintia, 80126 Napoli, Italy.}
\affiliation{Department of Physics E. Pancini, Università di Napoli Federico II, via Cintia, 80126 Napoli, Italy.}

\author{Vania Vellucci {\Large \orcidlink{0000-0003-4516-0940}}\,}
\email[]{vellucci@qtc.sdu.dk}
\affiliation{Quantum  Theory Center ($\hbar$QTC) \& D-IAS, Southern Denmark University, Campusvej 55, 5230 Odense M, Denmark.}

\begin{abstract}
\noindent
A variety of robust and effective descriptions have been devised to extract model-independent information about the fundamental properties of black holes from observational data when searching for deviations from general relativity. In this work, we construct explicit transformation maps establishing the equivalence 
among  three  relevant parametrizations for different spacetime patches: Johannsen–Psaltis, Rezzolla–Zhidenko, and Effective Metric Description.  We then select representative black hole geometries to determine the minimal number of parameters required within each scheme to reproduce the associated quasi-normal mode spectra with a prescribed degree of accuracy. Our analysis shows that, for the given observables, a finite set of coefficients suffices to attain the desired precision in the three frameworks. Finally, we emphasize how the individual strengths of these effective descriptions can be exploited to probe complementary aspects of black hole physics.
\end{abstract}

\maketitle

\section{Introduction}
\label{sec:intro}
Astrophysical black holes (BHs) constitute natural laboratories to test gravity in extreme regimes. They are among the most fascinating compact objects, as they are endowed with an event horizon hiding a region which is causally disconnected from the exterior. Although they may appear to be rather complex physical systems from an observational point of view, they are characterized only by their mass, charge, and spin, as predicted by general relativity (GR) \cite{Heusler:1996ft,Johannsen:2010gy}. This makes BHs remarkably simple objects despite their tangled origins. However, the fundamental laws governing them have not yet been robustly confirmed, requiring further investigations to encompass potential deviations from the classical GR metrics.

A strategy to explore such deformations was first introduced by Johannsen and Psaltis, incorporating a quadrupole moment independent of both mass and spin \cite{Johannsen:2010xs,Johannsen:2011dh}. This approach, based on previous developments \cite{Glampedakis:2005cf}, allows one to test the \emph{no-hair conjecture} via the examination of different physical phenomena, such as the location of the innermost stable circular orbit (ISCO) \cite{RotatingBHTeukolsky,Shapiro:1983du} or gravitational lensing effects \cite{Bozza:2010xqn} through the detection of potential modifications of the quadrupole moment. This model has been revised by Cardoso, Pani, and Rico \cite{Cardoso:2014rha} to address some of its weaknesses. Indeed, the original metric accounts only for corrections preserving the horizon area–mass relation. This limitation can be removed by introducing additional parameters, dominating in both weak- and strong-field regimes. The resulting framework yields the most general static and spherically symmetric BH geometry, characterized by twice the number of parameters of the original approach. This implies that, in the strong-field limit, all parameters contribute with comparable significance.

Building upon this idea, Rezzolla and Zhidenko proposed a parametrization for static and spherically symmetric BHs within metric theories of gravity \cite{Rezzolla:2014mua,Konoplya:2016jvv}. Their approach utilizes a continued-fraction expansion in terms of a compactified radial coordinate, offering better convergence properties compared to traditional Taylor series techniques. This method enables efficient approximations of various metrics with a reduced set of coefficients, with the hope of facilitating the comparison of observational data against predictions from different gravitational models (see e.g. \cite{Berti:2015itd,DeLaurentis:2017dny,Volkel:2020xlc,Suvorov:2021amy}).

More recently, in \cite{DelPiano:2023fiw,DelPiano:2024gvw,DamiaPaciarini:2024ssf,DamiaPaciarini:2025xjc} an independent alternative model-independent framework was developed to describe modifications of classical GR for BH metrics, termed the \emph{Effective Metric Description} (EMD). Here metric deformations are parametrized in terms of physical quantities, such as the radial proper distance. Focusing on static and spherically symmetric spacetimes, explicit expansions of the metric near the event horizon were constructed in terms of physical coefficients, allowing for the general, coordinate-independent parametrization of thermodynamic quantities, such as the Hawking temperature. In addition, the  asymptotic behavior  was analyzed and provided further  constraints on the metric. Altogether, these analyses lead to model-independent consistency conditions on metric deformations. The approach not only offers insights into the interplay between quantum effects and BH thermodynamics, or more generally on different sources of corrections to GR \cite{DAlise:2023hls,Binetti:2022xdi} but also in testing the mathematical consistency of generic BH models \cite{DelPiano:2023fiw,DelPiano:2024gvw,DamiaPaciarini:2024ssf,DamiaPaciarini:2025xjc} across different space-time dimensions. 

Collectively, these parametrizations serve to accommodate potential deviations from classical BH solutions that can help interpret GR precision tests. Ultimately, these frameworks should be used to bridge  the gap between alternative theories of gravity and  empirical observations. It is therefore desirable and timely to build a unified view of different agnostic parametrizations for deformed BH metrics. This is the overarching goal of this work. To this end, we will construct the transformation maps connecting the three aforementioned frameworks, highlighting the advantages and limitations of each approach. Establishing a coherent and concrete correspondence among the proposed parametrizations is crucial for a comparative analysis and a deeper understanding of their intrinsic features.

In recent years, BH parametrizations have also attracted remarkable attention in the field of BH perturbation theory \cite{Pani:2013pma,Sasaki:2003xr,Hui:2021cpm,Cano:2024jkd}. Once a BH is excited, it will reach an equilibrium configuration through the emission of gravitational waves. This relaxation process can be divided into three phases: an early response, depending on the initial conditions of the disturbance \cite{Chavda:2024awq}; a ringdown phase, where the fluctuation can be modeled as a superposition of complex frequencies called quasi-normal modes (QNMs) \cite{Konoplya:2011qq,Franzin:2023slm,Cardoso:2019mqo}; finally a late-time phase, where the perturbation decays as a power-law tail \cite{Okuzumi:2008ej,Mitsou:2010jv,Berti:2022xfj}. 
It turns out that modifications of the near-horizon metric can have non-negligible effects on the response of a BH to perturbations, especially in the QNM spectrum, which we will analyze in detail within the three aforementioned frameworks. However, the attention is focused on the so-called \emph{eikonal QNMs}, i.e., the limit of large angular momentum or multipole index $\ell \gg 1$~\cite{Dolan:2010wr, Cardoso2008}. There are several reasons behind the choice to investigate this observable in this limit. One is that, under suitable assumptions, the eikonal QNMs are directly related to the intrinsic properties of the underlying BH geometry \cite{Stefanov:2010xz}, such as the orbital frequency $\Omega$ (being also related to the BH shadow radius $b_{\rm sh}$ \cite{EventHorizonTelescope:2019dse,EventHorizonTelescope:2022wkp,EventHorizonTelescope:2022xqj,Perlick:2021aok}) and the Lyapunov exponent $\lambda$ of the photon sphere. The orbital frequency $\Omega$ and the Lyapunov exponent $\lambda$ (associated with the instability of photon sphere orbits) represent the real and imaginary parts of the QNM frequency, respectively. 
  
This work offers the opportunity to better understand the mathematical and physical link between the different parametrizations \cite{Cardoso:2014rha,Rezzolla:2014mua,DelPiano:2024nrl}, while simultaneously allowing a comparison between the QNM spectrum and the BH shadow results \cite{Konoplya:2020hyk}. 
 
The manuscript is structured as follows. In Section \ref{sec:comparison}, we introduce the various parametrizations. Specifically, we discuss the Johannsen-Psaltis (JP) \cite{Johannsen:2010xs} in subsection~\ref{sec:JP-parametrization}, the Rezzolla-Zhidenko (RZ) \cite{Rezzolla:2014mua} in \ref{sec:RZ-parametrization}, and the EMD \cite{DelPiano:2024gvw} in \ref{sec:EMD-parametrization}. In Section \ref{sec:parametrization-comparison} we provide the transformation maps relating the aforementioned parametrizations, where we also discuss their mathematical aspects and connections. Finally, in Sec. \ref{sec:eikonal_limit}, we consider the QNMs in the eikonal limit to compare the different parametrizations. Throughout this work, we adopt geometric units, namely $G=c=1$. We offer our conclusions in Sec.~\ref{sec:end} and in the appendices we provide helpful mathematical details for the EMD framework relevant for this work. 

\section{Agnostic parametrizations}
\label{sec:comparison}
Consider a generic static and spherically symmetric spacetime, whose line element in Schwarzschild coordinates $(t,r,\theta,\varphi)$ reads
\begin{equation}\label{eq:generic-metric}
\dd s^2=g_{tt}(r)\dd t^2+g_{rr}(r)\dd r^2+r^2\left(\dd\theta^2+\sin^2\theta\dd\varphi^2\right) \ ,
\end{equation}
where $g_{tt}$ and $g_{rr}$ are functions of the radial coordinate $r$ only. This is our starting point to introduce the Johannsen-Psaltis (JP) \cite{Johannsen:2010xs}, the Rezzolla-Zhidenko (RZ) \cite{Rezzolla:2014mua}, and the EMD \cite{DelPiano:2024gvw} parametrizations.  We further require the parametrizations to abide the experimental bounds existing on the deviations from GR in the weak-field regime, also known as parametrized post-Newtonian (PPN) constraints \cite{Will2014kxa,Williams:2004qba}.

\subsection{Johannsen-Psaltis parametrization}
\label{sec:JP-parametrization}
The JP parametrization originates from the need to test the no-hair conjecture with observations of BHs in the electromagnetic spectrum. This approach is expressed through a parametric spacetime containing a quadrupole moment independent of both mass and spin \cite{Johannsen:2010xs}. Therefore, any deviation from GR manifests in anomalous contributions to the quadrupole moment \cite{Glampedakis:2005cf,Johannsen:2010xs}.  However, this framework has been generalized by the authors in~\cite{Cardoso:2014rha} to overcome a series of issues. The original JP parametrization reads (cf. Eq.~\eqref{eq:generic-metric}):
\begin{equation}\label{eq:JP-metric}
g_{tt}^{\rm JP}=-[1+g(r)]\left(1-\frac{2\tilde{M}}{r}\right) \, , \quad g_{rr}^{\rm JP}=\frac{1+g(r)}{1-\frac{2\tilde{M}}{r}} \ ,
\end{equation}
where the function $g(r)$ is defined by the series:
\begin{equation} \label{eq:g-JP}
    g(r) \coloneqq \sum_{n=1}^\infty\epsilon_n \left(\frac{\tilde{M}}{r}\right)^n = \sum_{n=1}^\infty\frac{\epsilon_n}{2^n} \left(\frac{\hr}{r}\right)^n \ ,
\end{equation}
with $\epsilon_n$ real constants to be determined from observations, and $\tilde{M}$ is related to the horizon position $\hr=2\tilde{M}$ and to the gravitational mass-energy of the spacetime $M$ through:
%Arnowitt-Deser-Misner (ADM)
\begin{equation}
    M=\tilde{M}\left(1-\frac{\epsilon_1}{2}\right) \ .
\end{equation}
The revised JP parametrization becomes \cite{Cardoso:2014rha}
\begin{equation}\label{eq:revised-JP}
    g_{tt}^{\rm JP}= - \left[1+ g^t(r) \right]\left(1-\frac{2\tilde{M}}{r}\right) \, , \quad g_{rr}^{\rm JP}=\frac{1+g^r(r)}{1-\frac{2\tilde{M}}{r}} \ ,
\end{equation}
where the functions $g^t$ and $g^r$ are expanded as
\begin{equation} \label{eq:new g-JP}
    g^i(r) = \sum_{n=1}^\infty\epsilon_n^i \left(\frac{\tilde{M}}{r}\right)^n  \qq{with} i=t,r \ ,
\end{equation}
and the set of parameters $\{ \epsilon_i^t,\epsilon^r_i \}_{i\geq 1}$ has doubled with respect to the previous version of the JP parametrization. The comparison with the PPN parameters entails
\begin{subequations} \label{eq:relation-JPmass}
    \begin{align}
        M &= \tilde{M} \left( 1 - \frac{\epsilon_1^t}{2} \right)\ , \label{eq:mass}\\
        \epsilon_1^r &=  \gamma \,(2-\epsilon_1^t)-2 \ , \\
        2\epsilon_2^t &= (\beta - \gamma)(\epsilon_1^t-2)^2 + 4 \epsilon_1^t \ .
    \end{align}
\end{subequations}
Hence, the parameters $\epsilon_1^t$, $\epsilon_2^r$, and $\{ \epsilon_i^t,\epsilon^r_i \}_{i\geq 3}$ are not constrained, even in the case of GR, where $\beta=\gamma=1$.

\subsection{Rezzolla-Zhidenko parametrization}
\label{sec:RZ-parametrization}
The RZ parametrization aims to describe BH spacetimes in generic metric theories of gravity using a model-independent approach \cite{Rezzolla:2014mua}. The line element of the RZ parametrization reads (cf. Eq.~\eqref{eq:generic-metric})
\begin{align}\label{eq:RZ-metric}
g_{tt}^{\rm RZ}&=-N^2(r) \qq{and} g_{rr}^{\rm RZ}=\frac{B^2(r)}{N^2(r)} \ .
\end{align}
The BH event horizon is located at $r = \hr > 0$ and is defined by $N(\hr)=0$. The radial coordinate is then compactified by introducing the dimensionless variable $x := 1- \hr / r$, such that $x=0$ marks the location of the event horizon, while $x=1$ corresponds to spatial infinity.

We rewrite $N$ as $N^2=x A(x)$, where $A(x)>0$ for $0\leq x\leq1$. We further express the functions $A$ and $B$ in terms of the parameters $\epsilon$, $a_0$, and $b_0$, as follows:
\begin{subequations}\label{asympfix}
\begin{align}
\label{asympfix_1}
A(x)&=1-\epsilon (1-x)+(a_0-\epsilon)(1-x)^2+{\tilde A}(x)(1-x)^3 \ ,
\\
\label{asympfix_2}
B(x)&=1+b_0(1-x)+{\tilde B}(x)(1-x)^2 \ ,
\end{align}
\end{subequations}
where $\epsilon$ encodes deviations of $\hr$ from the Schwarzschild radius
$ \epsilon \coloneqq \frac{2M - \hr}{\hr}$ and the functions ${\tilde A}$ and ${\tilde B}$ describe the metric near the horizon (for $x \simeq 0$) and are finite there, as well as at spatial infinity (for $x \simeq 1$).

To achieve rapid convergence, these two functions are modeled via the Padé approximants in the form of continued fractions as
\begin{subequations}
\label{contfrac}
\begin{align}
\label{contfrac_1}
{\tilde A}(x)=\frac{a_1}{\displaystyle 1+\frac{\displaystyle
    a_2x}{\displaystyle 1+\frac{\displaystyle a_3x}{\displaystyle
      1+\ldots}}}\ ,\\
\label{contfrac_2}
{\tilde B}(x)=\frac{b_1}{\displaystyle 1+\frac{\displaystyle
    b_2x}{\displaystyle 1+\frac{\displaystyle b_3x}{\displaystyle
      1+\ldots}}}\ ,
\end{align}
\end{subequations}
where $a_1, a_2, a_3\ldots$ and $b_1, b_2, b_3\ldots$ are dimensionless constants determined from observations. 

Note that by comparing the large-distance expansion in Eq. \eqref{eq:RZ-metric} with the PPN one, we can put experimental bound on the first coefficients:
\begin{subequations}
\begin{align}\label{PPN bounds}
         a_0 &= \displaystyle \frac{(\beta - \gamma)(1+\epsilon)^2}{2} \lesssim10^{-5} \ ,\\
         b_0 &= \displaystyle \frac{(\gamma -1)(1+\epsilon)}{2} \lesssim 10^{-5} \ .
    \end{align}
\end{subequations}    
\subsection{Effective Metric Description parametrization}
\label{sec:EMD-parametrization}
The EMD parametrizes the deformations of the classical Schwarzschild metric in terms of spacetime invariants, which can be measured by observers independently from the set of coordinates, and preserve the same symmetries as GR. A natural choice for such a physical quantity is the radial proper distance to the BH horizon \cite{DelPiano:2023fiw,DelPiano:2024nrl,DelPiano:2024gvw}. The metric components can be written as
\begin{align}\label{eq:EMD-metric}
g_{tt}^{\rm EMD}&=-h(r) \qq{and} g_{rr}^{\rm EMD}=\frac{1}{f(r)} \ , 
\end{align}
where
\begin{equation}
    h(r) = 1 - \frac{\Psi(\mathcal{X})}{r} \qq{and} f(r) = 1 - \frac{\Phi(\mathcal{X})}{r} \ ,
\end{equation}
are positive definite for $r>r_{H}$ ($\hr$ being the BH event horizon) and the deformation functions $\Psi(\mathcal{X})$ and $\Phi(\mathcal{X})$ are parametrized by a physical quantity $\mathcal{X}$ that is monotonic in $r$ and invariant under coordinate reparametrizations. The specific choice of $\mathcal{X}$ does not affect the physical conclusions, since different parametrization schemes can always be locally mapped into each other \cite{DelPiano:2024gvw}.

We divide the discussion into regions close to (see Sec.~\ref{sec:EMD-near}) and far from (see Sec.~\ref{sec:EMD-large}) the BH horizon.

\subsubsection{EMD near the BH horizon}
\label{sec:EMD-near}
We consider $\Psi(\rho)$ and $\Phi(\rho)$, where $\rho=\rho(r)$, with $r \geq \hr$, is the radial proper distance measured from the BH horizon, which is fixed by the following differential equation
\begin{align}
%&\dv{\rho}{r}=\frac{1}{\sqrt{f(r)}} \qq{with} \rho(\hr)=0 \ 
    \dv{\rho}{r}=\left(1-\dfrac{\Phi(\rho)}{r}\right)^{-1/2} \qq{with} \rho(\hr)=0 \ .\label{SelfEquationDistance}
\end{align}
The solution can be expressed as a series expansion close to the BH horizon through
\begin{subequations}\label{SeriesDistanceRho}
\begin{align}
\Phi(\rho) &=\hr + 2M\sum_{n=1}^\infty \xi_n\,\left(\dfrac{\rho}{2M}\right)^n \ ,\label{SeriesDistanceRho_1}\\
\Psi(\rho) &=\hr + 2M\sum_{n=1}^\infty \theta_n\,\left(\dfrac{\rho}{2M}\right)^n\ \ ,\label{SeriesDistanceRho_2}
\end{align}
\end{subequations}
where $\xi_n$ and $\theta_n$ are real constants.\footnote{Let us remark that the constants introduced in this way are coinciding with the parameters $\mathfrak{x}_n$ and $\mathfrak{t}_n$ defined in \cite{DelPiano:2024nrl}.} The series~\eqref{SeriesDistanceRho} are assumed to have a non-vanishing radius of convergence, and all derivatives of $h$ and $f$ with respect to $r$ evaluated in $\hr$ are well defined. This implies $\xi_{2n-1}=\theta_{2n-1}=0$ for all $n\in\mathbb{N}$, and $\xi_2\leq \frac{M}{8 \hr}$ \cite{DelPiano:2024gvw}. We will adopt these assumptions for the rest of this work.

The proper distance $\rho(r)$ and its inverse $r(\rho)$ can be obtained via Eq.~\eqref{SeriesDistanceRho} as follows:
\begin{subequations}\label{SeriesDistance}
\begin{align}
r(\rho)&=\hr+2M \sum_{n=1}^\infty \mathfrak{a}_n \left(\frac{\rho}{2M}\right)^n\ ,\\
\rho(r)&=2M\sum_{n=1}^\infty \mathfrak{b}_n  \left(\frac{r-\hr}{2M} \right)^{n/2} \ ,
\end{align}
\end{subequations}
where the coefficients $\mathfrak{a}_n$ and $\mathfrak{b}_n$ can be found iteratively by solving the differential problem \eqref{SelfEquationDistance}. Explicit expressions for those coefficients are reported in \cite{DelPiano:2023fiw}.

The derivatives of the metric functions at the horizon can be written in terms of $\{\xi_{2n},\theta_{2n}\}_{n\geq1}$ as\footnote{Following the notation in \cite{DelPiano:2023fiw}, we denote the derivatives of the function $\phi$ with respect to the radial coordinate $r$ evaluated at the horizon $\hr$ as $\phi_{H}^{(n)} \coloneqq \dd^n \phi / \dd r^n \big|_{r=\hr}$.}
\begin{subequations}\label{first_order_expansion}
\begin{align}
   2 M \, \fHn{1}&=\frac{1+\sqrt{1-16 y_H \xi_2}}{2 y_H}\ , \label{f1h emd nh}\\ 
   %2 M \, h_H^{(1)}&=\frac{1-8 y_H \theta_2 + \sqrt{1-16 y_H \xi_2}}{y_H(1+\sqrt{1-16 y_H \xi_2})}
   2 M \, \hHn{1} &= \frac{1}{y_H}\left[1-\frac{\theta_2}{2 \xi_2}\left(1 - \sqrt{1-16 y_H \xi_2}\right) \right]\ , \label{h1h emd nh}
\end{align}
\end{subequations}
where $y_H \coloneqq \hr/(2M)$. The deformation to the higher-order derivatives take the form
\begin{equation}
    (2M)^{n}\left(\fHn{n}- (\fHn{n})_{\rm class} \right) \, \propto \,  \xi_{2n}+\mathrm{n.l.}(y_H,\xi_2,\ldots,\xi_{2n-2}) \ ,
\end{equation}
where $(\fHn{n})_{\rm class}$ is the classical expression for the derivative of the metric at the horizon (for Schwarzschild $y_H=1$ and $(\fHn{n})_{\rm class} = (-1)^{n+1}(2M)^{-n}$) and $\mathrm{n.l.}$ indicates a non-linear dependence on $y_H$ and $\{\xi_{2p}\}_{1 \leq p < n}$. The same holds for $\hHn{n}$ with respect to $\theta_{2n}$ and $(y_H,\{\theta_{2p}\}_{1 \leq p < n})$.

\subsubsection{EMD at large distance from the BH}
\label{sec:EMD-large}
We now consider the spacetime regions far from the BH horizon, where the spacetime is weakly curved, but still affected by the presence of the BH. Hence, we can asymptotically expand the deformation functions \cite{Binetti:2022xdi,DelPiano:2023fiw}. The full radial proper distance $d(r)$ from $r=0$ is then given by the differential problem
\begin{equation}\label{eqprop}
    \dv{d}{r}=\left|1-\dfrac{\Phi(d)}{r}\right|^{-1/2} \qq{with} d(0)=0 \ .
\end{equation}
In Appendix \ref{app:EMDlogs}, we show the steps to obtain the expressions of the metric functions as reported below
\begin{subequations}\label{eq:EMD-functions}
    \begin{align}
        f(r) &= 1 - \frac{2M}{r} - \frac{2M^2 \omega_1}{r^2} + (k \omega_1 - \omega_2)\frac{2M^3}{r^3}\notag\\
        &- \left[ \left(k^2  +  k   + \dfrac{3}{2}\right) \omega_1 + \omega_1^2 - 2 k \omega_2 +  \omega_3 \right]\frac{2M^4}{r^4}\notag\\
        &+ \order{\frac{M^5}{r^5}} \ , \\
        h(r) &= 1 - \frac{2M}{r} - \frac{2M^2 \gamma_1}{r^2} + (k \gamma_1 - \gamma_2)\frac{2M^3}{r^3}\notag+\\
        &- \left[ \left(k^2  +  k  + \dfrac{3}{2} \right) \gamma_1 + \gamma_1 \omega_1 - 2 k \gamma_2 +  \gamma_3 \right]\frac{2M^4}{r^4}\notag\\
        &+ \order{\frac{M^5}{r^5}} \ .
    \end{align}
\end{subequations}

It is important to emphasize that the asymptotic series in Eqs.~\eqref{eq:EMD-functions} must be interpreted with care. When truncating the parametrization to a finite number of deformation parameters, for instance $(\omega_1,\ldots,\omega_n)$ and $(\gamma_1,\ldots,\gamma_n)$, only the terms up to order $1/r^{n+1}$ should be retained. Coefficients at higher orders in $1/r$ are not reliable, because they implicitly depend on additional parameters that have been set to zero by the truncation. As a consequence, the apparent $1/r^{m}$ term with $m>n+1$ does not represent the true asymptotic behavior of the full parametrization, but rather a spurious artifact of the cutoff. This issue is absent in the Johannsen–Psaltis parametrization, where truncating the expansion automatically suppresses all higher-order terms, ensuring internal consistency. In contrast, in the EMD approach, truncation must be imposed manually by discarding all terms beyond the order supported by the retained number of parameters.

The combinations $k \omega_1-\omega_2$ and $k\gamma_1-\gamma_2$, as well as those appearing at higher orders, are independent of the choice of $k$. Indeed, observable quantities do not depend on the choice of this constant, see \cite{DAlise:2023hls,DelPiano:2024gvw} for details. We also note that when the expansion is truncated at $\order{M^4/r^4}$, the PN-expansions of $f$ and $h$ are equal when $\{\gamma_1, \gamma_2\}$ and $\{\omega_1, \omega_2\}$ are exchanged, whereas at the next leading order this is spoiled by the presence of mixed terms involving $\gamma_1 \omega_1$. 

By comparing \eqref{eq:EMD-functions} with the PPN expansion, we obtain:
\begin{align}\label{bounds}
    %|\beta-1|\lesssim 2.3\times10^{-4}\implies 
    |\gamma_1|=|\beta-\gamma|\lesssim 10^{-5}  \quad \text{with} \quad 
    \gamma= 1 \ .
\end{align}
At leading order in the PPN framework, we are able to map only $\gamma_1$ to $\beta$ and $\gamma$, while $\omega_1$ remains unconstrained. Conversely, the EMD expansion \eqref{eq:EMD-functions} is only able to capture a subset of PPN corrections, constrained by ${\gamma=1}$. 

\section{Transformation maps relating the different parametrizations}
\label{sec:parametrization-comparison}
In this section, we establish the relationships among the three distinct frameworks. Notably, while the event horizons in these parametrizations are defined through different formalisms, they all depend on free parameters. However, in all cases, they must uniquely identify the BH event horizon $\hr$. This observation is significant, as it facilitates the subsequent computational analyses. 

\subsubsection{JP to EMD (near horizon) parametrizations}
\label{sec:JP-EMD}
First, we compare the temporal metric coefficients $g_{tt}$ in the EMD \eqref{eq:EMD-metric} and JP \eqref{eq:revised-JP} parametrizations: 
\begin{align}\label{eq:h-g}
g_{tt}^{\rm EMD}&=-\sum_{n=1}^\infty \frac{h^{(n)}_H}{n!}(r-\hr)^n\notag\\
&=-\frac{1+g^t(r)}{r}(r-\hr)=g_{tt}^{\rm JP} \ .
\end{align}
Dividing both sides by $(r-\hr)$ we obtain
    \begin{align}\label{EMD_JP}
        &\sum_{n=1}^\infty \frac{h^{(n)}_H}{n!}(r-\hr)^{n-1}=\frac{1+g^t(r)}{r}= \notag\\
        &=\frac{1}{r}\left[1+\sum_{j=1}^\infty\frac{\epsilon_j^t}{2^j} \left(\frac{\hr}{r}\right)^j\right]\equiv  \sum_{j=0}^\infty\frac{\epsilon_j^t }{2^j} \frac{\hr^j}{r^{j+1}}\ ,
    \end{align}
where we included additional coefficient $\epsilon_0^t=1$ in the sum. Evaluating Eq. \eqref{EMD_JP} at $r=\hr$, we obtain for $n=1$
    \begin{equation} \label{eq:h1}
        h^{(1)}_H=\frac{1+g^t(\hr)}{\hr} = \frac{1}{\hr}\sum_{j=0}^\infty\frac{\epsilon_j^t}{2^j}  \ .
    \end{equation}
The coefficients $h^{(n)}_H$ can be computed by taking $n$ derivatives of Eq.~\eqref{EMD_JP} and then evaluating it at $r=\hr$. The general expression for the coefficients $h^{(n)}_H$ is thus:     
    \begin{align}
        h^{(n)}_H&=n \dv[n]{r} \left(\sum_{j=1}^{\infty}\frac{\epsilon_j^t}{2^j} \frac{\hr^j}{r^{j+1}}\right)\bigg|_{r=\hr}=\notag\\
    &=\frac{(-1)^n \, n}{\hr^{n+1}} \sum_{j=1}^{\infty}\frac{\epsilon_j^t }{2^j} \frac{(j+n)!}{j!}\ . \label{eq:hj}
    \end{align}
A similar approach can be applied to the radial metric component, yielding:
    \begin{align}
        g_{rr}^{\rm EMD}&=\frac{1}{f(r)}=\frac{1+g^r(r)}{1-\hr/r}=g_{rr}^{\rm JP}\ ,
     \end{align}   
where        
    \begin{align}
         f(r)&= \sum_{n=1}^\infty\frac{f^{(n)}_H}{n!}(r-\hr)^n \ .
    \end{align}
This can be also written as 
\begin{subequations} \label{eq:f_H}
    \begin{align}
        f^{(1)}_H&=\frac{1}{\hr[1+g^r(\hr)]} \ , \label{eq:f1} \\
        f^{(n)}_H&=n \dv[n]{r} \left[\frac{1}{r(1+g^r(r))}\right]\bigg|_{r=\hr} \ ,\label{eq:gr}
    \end{align}
\end{subequations}
where the general term $f_H^{(n)}$ cannot be easily found and then written in a closed form. 

Comparing Eqs.~\eqref{eq:f1} with \eqref{f1h emd nh}, Eqs.~\eqref{eq:h1} with \eqref{h1h emd nh}, and recalling that $\hr=2M/(1-\epsilon_1^t/2)$, we finally obtain
\begin{equation}
\xi_2 = \frac{g^r(\hr)(2-\epsilon_1^t)}{8[1+g^r(\hr)]^2} \ ,\quad
\theta_2 = \frac{g^t(\hr)(\epsilon_1^t-2)}{8[1+g^r(\hr)]} \ ,
\end{equation}
with $g^r(\hr) \neq -1$. The case $g^r(\hr) = 1$ saturates the bound on $\xi_2 \leq  (2-\epsilon_1^t)/32$. The expressions for the higher-order coefficients $\{\theta_{2n}, \xi_{2n} \}_{n\geq 2}$ are involved and we refrain from reporting them here. We note, however, that the higher-order derivatives of the metric functions depend linearly on these parameters, making their computation straightforward.

From these calculations we note that the EMD coefficients depend on an infinite number of JP terms, and vice versa. The JP coefficients can be written in terms of the EMD ones by inverting the linear system~\eqref{eq:hj}.

\subsubsection{JP to EMD (large distance) parametrizations}
The expansion of JP metric at large distance reads: 
\begin{subequations}\label{JPexpinf}
\begin{align}
    g_{tt}^{\rm JP}&=-1+\frac{\tilde{M}(2-\epsilon_1^t)}{r}+\frac{\tilde{M}^2(2\epsilon_1^t-\epsilon_2^t)}{r^2}+\order{\tilde{M}^3/r^3} \ ,\\
    g_{rr}^{\rm JP}&=1+\frac{\tilde{M}(2+\epsilon_1^r)}{r}+\frac{\tilde{M}^2(\epsilon_2^r+2\epsilon_1^r+4)}{r^2}+\order{\tilde{M}^3/r^3} \ .
\end{align}
\end{subequations}

%{\bf JP to EMD}:
Comparing \eqref{JPexpinf} with the EMD expansion \eqref{eq:EMD-functions}, and taking into account the relation \eqref{eq:mass}, we finally obtain:
\begin{subequations}\label{JP to EMD fr}
    \begin{align}
        \epsilon_1^r & = - \epsilon_1^t \ , \\ 
        \omega_1 &= \frac{2[\epsilon_2^r+ \epsilon_1^t(2-\epsilon_1^t)]}{(\epsilon_1^t - 2)^2},\ \ \gamma_1 = \frac{2(2\epsilon_1^t-\epsilon_2^t)}{(\epsilon_1^t - 2)^2}, \\
        \omega_2 & =-\frac{2}{(\epsilon_1^t - 2)^3}\left\{2\epsilon_3^r+[k(\epsilon_1^t-2)-2\epsilon_1^t](\epsilon_1^t-2)\epsilon_1^t \right. \notag \\ & \qquad \left.+\epsilon_2^r[4(\epsilon_1^t-1)+k(2-\epsilon_1^t)]\right\} \ , \\
        \gamma_2 &= \frac{2k(\epsilon_1^t-2)(2 \epsilon_1^t-\epsilon_2^t)-8\epsilon_2^t+4\epsilon_3^t}{(\epsilon_1^t - 2)^3} \ .    \end{align}
\end{subequations}
where we write only the first few terms. 

We note that this comparison can be easily carried out, because there is a one-to-one correspondence among the JP and EMD coefficients. This is due to the asymptotic expansion of the JP metric, which establishes a direct link among the two parametrizations. 

\subsubsection{JP to RZ parametrizations}
\label{sec:JP-RZ}
We note that the map between Eqs. \eqref{eq:RZ-metric} and \eqref{eq:JP-metric} has already been presented in Ref. \cite{Rezzolla:2014mua}. Here, we adapt it to the revised JP parametrization \eqref{eq:revised-JP}. Comparing that with Eqs.~\eqref{asympfix_1} and \eqref{asympfix_2}, we have
\begin{subequations}
    \begin{gather}
        -\epsilon \frac{\hr}{r}+(a_0-\epsilon)\left(\frac{\hr}{r}\right)^2 + \left(\frac{\hr}{r}\right)^3\tilde{A}\left(1-\frac{\hr}{r}\right) = g^t(r) \ , \\
        b_0 \frac{\hr}{r} + \left(\frac{\hr}{r}\right)^2\tilde{B}\left(1-\frac{\hr}{r}\right) =  g^{B}(r)\ .
    \end{gather}
\end{subequations}
where we introduced the auxiliary function 
\begin{equation}
g^{B}(r)\coloneqq\sqrt{[1+g^t(r)][1+g^r(r)]}-1 \ , %= \sqrt{1 + \sum_{i=1}^\infty \left[\epsilon_i^t+\epsilon_i^r + \sum_{j=1}^i \epsilon_j^t \epsilon_{i-j}^r \right]\frac{(1-x)^i}{2^i}}-1\ ,    
\end{equation}
The first few coefficients can be extracted as
\begin{equation}
\label{JPrel}
\epsilon =-\frac{\epsilon_1^t}{2}\ , \ a_0=\frac{1}{2}\left( \frac{\epsilon_2^t}{2} - \epsilon_1^t\right)\ , \ b_0=\frac{\epsilon_1^t+\epsilon_1^r}{4}\ .
\end{equation}
By matching the two parametrizations near the horizon, we obtain algebraic relations between the RZ coefficients $\{a_n, b_n\}_{n \geq 1}$ and the JP coefficients $\{\epsilon_n^t, \epsilon_n^r \}_{n \geq 1}$. The first few expressions are displayed below:
\begin{subequations}\label{as_bs_JP}
\begin{align}
a_1&=g^t(\hr)- a_0 + 2 \epsilon =\sum_{n=3}^{\infty}\frac{\epsilon_n^t}{2^n}\ ,\\ 
b_1&= g^B(\hr)-b_0 \ ,\\
a_2&= - \frac{(r\, g^t)^{\prime}+ g^t+\epsilon}{a_1}\Big|_{r=\hr}-1\notag \\
& =\displaystyle\sum_{n=4}^{\infty}\frac{\epsilon^t_n(n-3)}{2^n} \Big/\displaystyle\sum_{n=3}^{\infty}\frac{\epsilon^t_n}{2^n}\ , \\
b_2&=- \dfrac{(r \, g^B)^{\prime}}{b_1}\bcond -1 \ ,\label{a2_b2_JP}  \\
a_3&=\frac{(r^2 g^t)^{\prime \prime}-2a_1[1+a_2(a_2+2)]}{2 a_1 a_2}\Big|_{r=\hr}\ ,\\
b_3&=\dfrac{(r^2 g^B)''}{2b_1 b_2}\bcond - (b_2 + 1)  \ ,\label{a3_b3_JP}
\end{align}
\end{subequations}
where the prime indicates the derivative with respect to $r$. Naturally, in Eq.~\eqref{as_bs_JP} we can easily extend to higher orders if needed. 

Finally, we remark that the terms $a_1,a_2,a_3\ldots$ do not depend on $\epsilon_1^t$ and $\epsilon_2^t$, whereas the terms $b_1,b_2,b_3\ldots$ follow a more complicated expression, which we chose not to display here. In the simplest case when $\epsilon_3^t\neq0$ and $\epsilon^t_n=0$ for $n>3$, we have $a_2=0$ and the approximant for the function $N(r)$ reproduces it exactly. This illustrates how the RZ coefficients can be related to the JP ones. The inverse map can be achieved by inverting the linear system \eqref{as_bs_JP}. 

\subsubsection{RZ to EMD (near horizon) parametrizations}
\label{sec:RZ-EMD}
The final map to consider is between the RZ and EMD parametrizations. Throughout the calculations we make use of the results from Sec. \ref{sec:JP-RZ}. By employing Eqs. \eqref{eq:h-g} and \eqref{eq:gr}, we obtain the following two relations:
\begin{subequations}
\begin{align}
g^t(r)&=r\left[\sum_{n=1}^\infty\frac{h_H^{(n)}}{n!}(r-\hr)^{n-1}\right]-1\ ,\\
g^r(r)&=\dfrac{1}{\sum_{n=1}^\infty\frac{f^{(n)}_H}{n!}(r-\hr)^{n-1}}-1\ .
\end{align}
\end{subequations}
We can now exploit Eq. \eqref{as_bs_JP} to express the RZ coefficients in terms of the derivatives of $g^t(r)$ and $g^r(r)$, which, in turn, are related to the EMD coefficients.

We can obtain the direct link between the first-order coefficients $(\theta_2,\xi_2)$ and $(a_0,a_1,b_0,b_1)$ of the EMD and RZ parametrizations, respectively, by using $h^{(1)}_H$ and $f^{(1)}_H$ (cf. Eqs. \eqref{eq:h1} and \eqref{eq:f1}) and
\begin{equation}
    h(r) = N(r)^2 \qq{and} f(r)=\frac{N(r)^2}{B(r)^2} \ .
\end{equation}
In order to determine the relation between the coefficients, let us first define $y=r/(2M)$ and then expand the RZ metric functions up to the first order in $y-y_H$:
\begin{subequations}\label{N2 and N2B2 series}
\begin{align}
    N(y)^2 &= \frac{(3+a_0+a_1)y_H - 2}{y_H^2} (y-y_H) + \mathcal{O}(y-y_H)^2 \ ,  \\
    \frac{N(y)^2}{B(y)^2} &= \frac{(3+a_0+a_1)y_H - 2}{(1+b_0+b_1)^2 \, y_H^2} (y-y_H) + \mathcal{O}(y-y_H)^2  \ .
\end{align}
\end{subequations}
Comparing these expansions with the first derivatives of Eq. \eqref{first_order_expansion} and solving for $\xi_2$ and $\theta_2$, we obtain
\begin{subequations}\label{RZtoxitheta}
    \begin{align}
        \xi_2 &= \frac{(1+\epsilon)(1+\mathcal{A}-2\epsilon)(\mathcal{B}(\mathcal{B}+2)-\mathcal{A}+2\epsilon)}{4(1+\mathcal{B})^4}\ , \label{RZtoxitheta-1}\\
        \theta_2 &= \frac{(1+\epsilon)(2\epsilon-\mathcal{A})}{8} \left( 1 + \frac{| 1+ 2\mathcal{A} - \mathcal{B}(\mathcal{B}+2)  - 4 \epsilon| }{(1+\mathcal{B})^2} \right) \label{RZtoxitheta-2} \ ,    
    \end{align}
\end{subequations}    
where $\mathcal{A} \coloneqq a_0+a_1$ and $\mathcal{B} \coloneqq b_0+b_1$. It is worth noticing that the parameters $\xi_2$ and $\theta_2$ depend on the RZ parameters only through the combinations $\mathcal{A}$ and $\mathcal{B}$. Moreover, the coefficients $a_n$ and $b_n$ with $n > 1$ do not contribute to the first-order parameters $\xi_2$ and $\theta_2$. 

Additional constraints on the parameters can be obtained by observing that, in the limit $b_0 = b_1 = 0$, we have $\mathcal{B}=0$. At the horizon, $B(y) \equiv 1 + \order{(y-y_H)^2}$, so the first derivatives at the horizon in Eqs.~\eqref{N2 and N2B2 series} must coincide with~\eqref{f1h emd nh} and~\eqref{h1h emd nh}. This implies the additional condition $\xi_2 = \theta_2$. Furthermore, when setting $\mathcal{B}=0$, the absolute value in Eq.~\eqref{RZtoxitheta-2} must be taken into account, which requires imposing 

\begin{equation}
    1+2\mathcal{A} - 4 \epsilon > 0  \ \Rightarrow \ 2 \epsilon - a_0 - a_1 < \frac{1}{2}\ .
\end{equation}

The relations in Eq.~\eqref{RZtoxitheta} can be inverted to obtain the coefficients $\mathcal{A}$ and $\mathcal{B}$ as functions of $\theta_2$ and $\xi_2$. As previously noted, if $\theta_2=\xi_2$, then the first derivatives of $f(y)$ and $h(y)$ at $y_H$ must coincide. Moreover, in the limit $\epsilon\to 0$ (see under Eq.~\eqref{asympfix}) the EMD and RZ parametrizations should recover the Schwarzschild BH. Therefore, we have:
        \begin{subequations}
        \begin{align}
            \mathcal{A}&=\dfrac{1}{2\sqrt{2}\,\xi_2}\Big{[}8 \theta_2 \xi_2 + (1+\epsilon)(\xi_2- \theta_2)\notag\\
            &\quad+\sqrt{1+\epsilon}\, (\xi_2-\theta_2) \sqrt{1+\epsilon-16 \xi_2}\Big{]}^{1/2}-1 \ ,\\
            \mathcal{B}&=2 \epsilon -\frac{\theta_2}{2 \xi_2}\left(1- \sqrt{1-\frac{16\xi_2}{1+\epsilon}}\right) \ .
        \end{align}            
        \end{subequations}
This approach can be straightforwardly generalized to higher-order coefficients by following the procedure already described for the first-order terms. The expressions beyond first order become rather cumbersome, and hence we choose not to display them here. We conclude there exists a well-defined and direct correspondence between the coefficients of the EMD and RZ parametrizations.

\subsubsection{RZ to EMD (large distance) parametrization}
The comparison between RZ and EMD parametrizations at large distance can be obtained by considering the following expansion for the RZ metric components:
\begin{subequations}
\begin{align}
g_{tt}^{\rm RZ}&=-1+\frac{2 M}{r}-\frac{4 a_0 M^2}{ (\epsilon +1)^2 r^2}\notag\\
&\qquad -\frac{8M^3 \left[\tilde{A}(1)-a_0 + \epsilon \right]}{ (\epsilon +1)^3r^3}+\mathcal{O}\left(\frac{M^4}{r^4}\right) \ ,\\
g_{rr}^{\rm RZ}&=1+\frac{2 M(2 b_0 + \epsilon +1)}{ (\epsilon +1)r}+\mathcal{O}\left(\frac{M^2}{r^2}\right) \ .
\end{align}    
\end{subequations}
Comparing these expansions with Eq. \eqref{eq:EMD-functions}, we obtain $b_0=0$ and the following conditions on the first few parameters:
\begin{subequations}\label{RZ to EMD large distance}
    \begin{align}
        \omega_1 & =\frac{ 4 \tilde{B}(1)- 2 a_0}{(\epsilon +1)^2}, \quad \gamma_1 = -\frac{2 a_0}{(\epsilon +1)^2} \ , \\
        \gamma_2 &= -\frac{2 \left[a_0 (k(1+\epsilon) -2)+2 (\tilde{A}(1)+\epsilon )\right]}{(\epsilon +1)^3} \ , \\
        \omega_2 & =-\frac{2}{(\epsilon +1)^3} \Big{\{} a_0 [k (1+\epsilon)-2] +2 \Big{[}\epsilon + \tilde{A}(1)\notag \\
        &  + (2-k)(1+ \epsilon)\tilde{B}(1) + 2 \tilde{B}^\prime(1)\Big{]} \Big{\}} \ ,
    \end{align}
\end{subequations}
where we note that the full continued fractions $\tilde{A}(1)$ and $\tilde{B}(1)$ (cf. Eq.~\eqref{contfrac}) and their derivatives appear, with the prime denoting differentiation with respect to $x$ defined in Sec.~\ref{sec:RZ-parametrization}.
As mentioned previously, the coefficients of RZ and EMD parametrizations at large distance are in one-to-one correspondence.

\section{Eikonal limit}
\label{sec:eikonal_limit}
As an application of the mapping among the three parametrizations, we now consider the eikonal limit of QNMs within the framework of BH perturbation theory. 

QNMs are the characteristic complex oscillation modes of perturbed compact objects \cite{Berti2009}; their real and imaginary parts correspond to the oscillation frequency and the inverse damping time, respectively. For a static and spherically symmetric BH metric~\eqref{eq:generic-metric}, the QNMs $\omega$ satisfy the following Schrödinger-like equation:
    \begin{equation}
        \dv[2]{\Psi}{r_*}+(\omega^2-V(r))\Psi=0\ ,
    \end{equation}
with outgoing boundary conditions $e^{\pm i\omega r_*}$ at the BH event horizon and at spatial infinity, expressed in terms of the tortoise coordinate $r_\ast$, defined through $\dd r_*/ \dd r = \sqrt{-g_{rr}/g_{tt}}$. Here, $V(r)$ denotes the perturbation potential, which depends on the spin of the perturbing field $\Psi$. Within the QNM spectrum, a notable role is played by the eikonal limit, corresponding to the regime of very large angular momentum, or multipole index, $\ell \gg 1$. In this limit, the QNM frequency reads
\begin{equation}\label{eq:eikonal}
        \omega_{n \ell}=\Omega \,\ell-i\left(n+\frac{1}{2}\right)|\lambda|\ ,
\end{equation}
where $n \in \mathbb{N}$ is the overtone number, $\Omega$ is the orbital frequency of light rays at the photon-sphere radius $r_{\rm ps}$, and $\lambda$ is the Lyapunov exponent \cite{Cardoso2008}.

The eikonal limit establishes a correspondence between the QNMs and the properties of the photon sphere, under the following two criteria \cite{Konoplya2017}: (1) the perturbation potential is positive definite (to avoid instabilities), single peaked, and decays to zero at the boundaries; and (2) the perturbation is a test scalar field or other field minimally coupled to gravity. This implies that the eikonal QNMs associated with gravitational perturbations may not, in general, be directly related to the properties of the unstable photon orbit. For this reason, we will restrict our analysis to perturbations of test scalar and electromagnetic fields. 

Given a static and spherically symmetric spacetime in Eq. \eqref{eq:generic-metric}, the photon dynamics can be described using the effective potential $U^2(r)=-g_{tt}(r)/r^2$. The radius, orbital frequency \cite{Bardeen:1972fi}, and Lyapunov exponent of the photon sphere are then given by \cite{Cardoso2008}\footnote{Note the change of the metric signature with respect to Eqs.~(35),~(37), and~(40) reported in Ref.~\cite{Cardoso2008}.} 
\begin{subequations}\label{eq:EIKONAL-QUANTITIES}
\begin{align}
&\rps \, g_{tt}^\prime(\rps)=2 \, g_{tt}(\rps) \ ,\\
\Omega^2 & = U^2(r_{\rm ps})= -\frac{g_{tt}(r)}{r^2}\bigg|_{r=\rps} \ , \label{eq:orbital_frequency}\\
\lambda^2 & = -\frac{\rps^2}{2  \, g_{rr}(\rps)}\dv[2]{U^2(r)}{r}\bigg|_{r=\rps} \ ,\label{eq:Lyapunov_exponent}
\end{align}          
\end{subequations} 
respectively. We note that $\Omega$ is closely related to the radius of the BH shadow through $\bsh :=U(\rps)^{-1}\equiv \Omega^{-1}$ \cite{Perlick:2021aok}.\footnote{In Ref. \cite{Cardoso2008}, the authors use a different potential $V_r(r)$ instead of $U^2(r)$.}

\subsection{Eikonal limit in different parametrizations}
In this subsection, we investigate the eikonal limit within the three BH parametrizations introduced in Sec.~\ref{sec:comparison}. For a generic parametrization, the corresponding set of infinite coefficients is treated as independent. However, once a specific model is considered, these coefficients become functions of a finite number of free parameters.

In the next subsection, we examine small deviations from the Schwarzschild solution using models specified by the mass $M$ and a single additional  $l$, which quantifies the departure from the GR geometry. This introduces a displacement of the BH horizon radius $\hr$ from its Schwarzschild value $2M$. To describe this effect, we adopt the perturbative parameter $\epsilon$ (see below Eq.~\eqref{asympfix}), which allows for a Taylor expansion of the quantities that enter the eikonal QNM limit. Other choices, such as $\epsilon_1^t$ in the JP parametrization, are equally possible.

We then expand the physically relevant quantities in Eq.~\eqref{eq:EIKONAL-QUANTITIES} in the eikonal limit up to linear order in $\epsilon$ for each parametrization. Higher-order terms in $\epsilon$ are subsequently used to estimate the truncation error.

\subsubsection{Rezzolla-Zhidenko parametrization}
\label{EL:RZ}
We start with the RZ parametrization, where the $\epsilon$ parameter emerges naturally. We assume the coefficients $a_n$ and $b_n$ admit a power-series expansion in $\epsilon$ of the form
\begin{equation}\label{RZ epsilon scaling}
    a_n(\epsilon) = \sum_{p=0} A_{n,p} \, \epsilon^p \qq{and} b_n(\epsilon) = \sum_{p=0} B_{n,p}\, \epsilon^p \ ,
\end{equation}
for $n\geq 0$ and $A_{0,0} = A_{1,0} = B_{0,0} = B_{1,0} = 0$. The last condition ensures that the leading contributions to the metric functions from $a_0(\epsilon)$, $a_1(\epsilon)$, $b_0(\epsilon)$, and $b_1(\epsilon)$ vanish in the Schwarzschild limit $\epsilon \to 0$, thereby guaranteeing that the deformations associated with Eqs.~\eqref{asympfix_1} and \eqref{asympfix_2} are continuously switched off.

The values of the coefficients $A_{n,p}$ and $B_{n,p}$ depend on the specific model under consideration. In general, these coefficients appear in the series expansion of the quantities in Eqs.~\eqref{eq:EIKONAL-QUANTITIES} through nonlinear combinations at each order in $\epsilon$. These expressions simplify dramatically when $A_{2,0} = B_{2,0} = 0$:
\begin{subequations}\label{eikonal q rz}
    \begin{align}
        \frac{\rps^{\rm RZ}}{M} &= 3 - \frac{4}{9}\left( 5+A_{0,1}+A_{1,1} \right)\, \epsilon + \order{\epsilon^2} \ , \\
        \Omega^{\rm RZ} M &=  \frac{1}{3 \sqrt{3}} + \frac{2}{81\sqrt{3}} \left( 6 +3 A_{0,1}+2A_{1,1} \right) \, \epsilon + \order{\epsilon^2}\  , \\
        \lambda^{\rm RZ} M &= \frac{1}{3 \sqrt{3}} + \frac{2}{81\sqrt{3}} \left( -4 + 7 A_{0,1}+ 4 A_{1,1}  \right. \notag \\
        & \qquad \left. + 9 B_{0,1} - 6B_{1,1}\right) \, \epsilon + \order{\epsilon^2} \ . 
    \end{align}
\end{subequations}
In the following sections, we restrict our attention to BH metrics satisfying the aforementioned condition.

\subsubsection{Johannsen-Psaltis parametrization}
\label{EL:JP}
Following the same reasoning as for the RZ parametrization, we note that the deformation functions $g^t(r)$ and $g^r(r)$ must vanish in the Schwarzschild limit. This implies that all JP parameters $\{ \epsilon_n^t , \epsilon_n^r \}_{n\geq 1}$ scale homogeneously with $\epsilon$. Indeed, as shown in Eq.~\eqref{JPrel}, we know that $\epsilon_1^t = -2\, \epsilon$ exactly. Therefore, for $\epsilon \to 0$ and the scaling behavior of Eq.~\eqref{RZ epsilon scaling}, we generally have $\epsilon_1^r , \epsilon_2^t \sim \order{\epsilon^k}$, with $k \geq 1$. Hence, we assume the following general power series expansion
\begin{subequations}\label{JP epsilon scaling}
    \begin{align}
        \epsilon_n^t(\epsilon) = \sum_{p=1}(\epsilon^t_n)_p \, \epsilon^p \qq{for} n \geq 2 \ ,\\
        \epsilon_n^r(\epsilon) = \sum_{p=1}(\epsilon^r_n)_p \, \epsilon^p \qq{for} n \geq 1 \ ,
    \end{align}
\end{subequations}
which could be equivalently read as a power series in $\epsilon_1^t$. The expressions for the coefficients $(\epsilon^t_n)_p$ and $(\epsilon^r_n)_p$ are determined by the model under investigation. Inserting the expansions~\eqref{JP epsilon scaling} into Eq.~\eqref{eq:EIKONAL-QUANTITIES} written in the JP parametrization (truncated for brevity at $\epsilon_4^t$ and $\epsilon_4^r$), we obtain 
\begin{subequations}\label{eikonal q jp}
    \begin{align}
        \frac{\rps^{\rm JP}}{M} &= 3 - \left( \frac{8}{3} + \frac{(\epsilon_2^t)_1}{9}+\frac{(\epsilon_3^t)_1}{18} + \frac{2(\epsilon_4^t)_1}{81} \right) \epsilon + \order{\epsilon^2} \ , \\
        \Omega^{\rm JP} M &= \frac{1}{3\sqrt{3}} + \bigg( \frac{2}{9} + \frac{(\epsilon_2^t)_1}{54} +\frac{(\epsilon_3^t)_1}{162}  + \frac{2(\epsilon_4^t)_1}{486} \bigg)\frac{ \epsilon}{\sqrt{3}} + \order{\epsilon^2}\  , \\
        \lambda^{\rm JP} M &= \frac{1}{3\sqrt{3}} + \bigg( \frac{5}{27} + \frac{2(\epsilon_2^t)_1}{81} + \frac{(\epsilon_3^t)_1}{162} + \frac{(\epsilon_4^t)_1}{1458}  \notag \\
        &\quad - \frac{(\epsilon_1^r)_1}{18} - \frac{(\epsilon_2^r)_1}{54} - \frac{(\epsilon_3^r)_1}{162}- \frac{(\epsilon_4^r)_1}{486} \bigg)\frac{ \epsilon}{\sqrt{3}}+ \order{\epsilon^2} \ . 
    \end{align}
\end{subequations}
As expected from Eqs. \eqref{as_bs_JP} and \eqref{eikonal q rz}, all JP parameters enter linearly at the first order in $\epsilon$.

\subsubsection{EMD (near horizon) parametrization}
\label{EL:EMD-near}
As shown in~\cite{DelPiano:2024nrl}, the EMD parametrization near the horizon can be extended up to the photon sphere by employing the Padé approximation~\cite{bender78:AMM} (see Appendix~\ref{app:emd eikonal} for details). In this case, the position of the event horizon is written as
\begin{equation}\label{r_horizonEMD}
    \hr = 2M (1 + \mathfrak{c}) \ ,
\end{equation}
and hence, the expressions for the quantities reported in Eq.~\eqref{eq:EIKONAL-QUANTITIES} can be compactly written as
\begin{subequations}\label{eikonal q emd nh}
    \begin{align}
        \frac{\rps^{\rm EMD(nh)}}{M} &= \sigma_0 + \sigma_1 \, \mathfrak{c} + \order{\mathfrak{c}^2}\ , \\
        \Omega^{\rm EMD(nh)} M &= \eta_0 + \eta_1 \, \mathfrak{c} + \order{\mathfrak{c}^2} \  , \\
        \lambda^{\rm EMD(nh)} M &= \Upsilon_0 + \Upsilon_1 \, \mathfrak{c} + \order{\mathfrak{c}^2} \ . 
    \end{align}
\end{subequations}
where $\mathfrak{c}$ quantifies the departure of $\hr$ from the Schwarzschild radius and is related to the RZ parameter as
\begin{equation}\label{c_EMD}
    \mathfrak{c}=-\frac{\epsilon}{\epsilon+1} \ .
\end{equation}
The coefficients are linear combinations of the parameters $\{\theta_{2n},\xi_{2n}\}_{n \geq 1}$ and their explicit form depends on the order of the employed Padé approximant, see Table I in~\cite{DelPiano:2024nrl} and Table~\ref{Tab:potential_Padé} in Appendix \ref{app:emd eikonal}. A Padé approximation of order $(N,M)$ involves a total of $N+M$ parameters. We note that the effective parameters $\{\theta_{2n}, \xi_{2n}\}_{n \geq 1}$ must still be expanded in power series of $\epsilon$, as we will see in the subsection dedicated to specific BH models.
\subsubsection{EMD (at large distance) parametrization}
\label{EL:EMD-large}
The EMD parametrization at large distances must recover the Schwarzschild metric when the deformation parameters $\{\gamma_i, \omega_i \}_{i\geq 1}$ vanish. However, the position of the event horizon does not explicitly appear as an input, and therefore there is no direct connection to the $\epsilon$ parameter. Nevertheless, by inspecting the relations Eqs.~\eqref{RZ to EMD large distance} and ansatz~\eqref{RZ epsilon scaling} (or equivalently by considering Eq.~\eqref{JP to EMD fr} and ansatz~\eqref{JP epsilon scaling}), we find that $\{\gamma_i, \omega_i \}_{i\geq 1}$ indeed admit a power series expansion in $\epsilon$.

Therefore, we assume that $\{\gamma_i, \omega_i \}_{i\geq 1}$ can be expanded similarly to the previous cases, namely
\begin{equation}
    \gamma_n (\epsilon) = \sum_{p=1} \gamma_{n,p} \, \epsilon^p \qq{and} \omega_n (\epsilon) = \sum_{p=1} \omega_{n,p} \, \epsilon^p \ ,
\end{equation}
where $\{\gamma_{n,p},\omega_{n,p}\}_{n\geq 1,\ p \geq 1}$ assume different values depending on the model under consideration.
Using the metric functions in Eqs.~\eqref{eq:EMD-functions}, we can express Eqs.~\eqref{eq:EIKONAL-QUANTITIES} as
\begin{subequations}\label{eikonal q emd ld}
    \begin{align}
        \frac{\rps^{\rm EMD(ld)}}{M} &= 3 + \frac{1}{9}\Big[(15-3k+2k^2) \gamma_{1,1}  \notag \\
        & \quad + (5-4k)\gamma_{2,1}+2\gamma_{3,1} \Big] \, \epsilon + \order{\epsilon^2}\ , \\
        \Omega^{\rm EMD(ld)} M &= \frac{1}{3 \sqrt{3}} + \frac{1}{162\sqrt{3}} \Big[ (-2 k^2 + 4 k - 21) \gamma_{1,1}  \notag \\
        &\quad + (4k-6) \gamma_{2,2} - 2 \gamma_{3,1}\Big] \, \epsilon + \order{\epsilon^2} \  , \\
        \lambda^{\rm EMD(ld)} M &= \frac{1}{3 \sqrt{3}} + \frac{1}{162\sqrt{3}} \Big[ (6 k^2 - 4 k + 21) \gamma_{1,1}  \notag    \\
        &+ 2 (5k-6) \gamma_{2,2} + 6 \gamma_{3,1} - (2k^2 - 4k + 21) \omega_{1,1} \notag  \\
        &+ (4k -6) \omega_{2,1} - 2 \omega_{3,1}\Big] \, \epsilon + \order{\epsilon^2} \  , 
    \end{align}
\end{subequations}
Above we kept the dependence on the gauge parameter $k$ to keep the expressions as general as possible.

\subsection{Application to BH models}
As an application of the formalism developed in the previous subsection, let us consider the following BH spacetimes: Hayward~\cite{Hayward_2006}, Bardeen~\cite{Bardeen1968}, and Simpson-Visser II~\cite{Simpson2019}. These are regular BH candidates%\footnote{Regular BHs are models that share the properties of standard GR BHs, except that they do not possess an essential singularity at the center.}
, which can be effectively described as deformations of the Schwarzschild geometry and satisfy the condition $g_{tt} \, g_{rr}=-1$. Their $g_{tt}$ metric components read

\begin{subequations}
    \begin{align}
        -g_{tt}^{\rm H}(r) &= 1 - \frac{2M r^2}{r^3 + 2 M \,l_{\rm H}^2} \ , \label{gtt Hayward} \\
        -g_{tt}^{\rm B}(r) &= 1 - \frac{2M r^2}{(r^2 + l_{\rm B}^2)^{3/2}} \ , \label{gtt Bardeen}\\
        -g_{tt}^{\rm SV2}(r) &= 1 - \frac{2M}{r}\exp{-\frac{l_{\rm SV2}}{r}} \ . \label{gtt SV2}
    \end{align}
\end{subequations}
The regularization length-scale parameters $(l_{\rm H},l_{\rm B},l_{\rm SV2})$ have different physical interpretations in each model, are allowed to vary over a finite range, and can be expressed in terms of $\epsilon$ as follows:
\begin{subequations}
    \begin{align}
        l_{\rm H} &= 2M \sqrt{\frac{\epsilon}{(1+\epsilon)^3}} \ , \\
        l_{\rm B} &= 2M \frac{\sqrt{(1+\epsilon)^{2/3}-1}}{1+\epsilon}  \ , \\
        l_{\rm SV2} &= 2M \frac{\log(1+\epsilon)}{1+\epsilon} \ .
    \end{align}
\end{subequations}

\begin{table*}[ht]
    \centering
    \renewcommand{\arraystretch}{2.5} % increase vertical spacing between rows
    
    \begin{tabular}{|wc{2.8cm}|wc{3.3cm}|wc{2.5cm}|wc{3.2cm}|wc{3.7cm}|}
        \hline
        \raisebox{0.07cm}{BH models} & \raisebox{0.07cm}{Parametrizations} & \raisebox{0.12cm}{$\dfrac{\rps}{M} - 3$} & \raisebox{0.12cm}{$\Omega M - \dfrac{1}{3\sqrt{3}}$} & \raisebox{0.12cm}{$\lambda M - \dfrac{1}{3\sqrt{3}}$} \\ 
        \hline
        \multirow{5}{*}{\raisebox{-0.22cm}{Hayward}}
        & From the model & $-\dfrac{16}{9}\epsilon + \dfrac{80}{27}\epsilon^2 $ 
                & \raisebox{0.015cm}{$\dfrac{8}{81 \sqrt{3}}\epsilon - \dfrac{40}{243 \sqrt{3}}\epsilon^2 $} 
                & \raisebox{0.015cm}{$-\dfrac{16}{81 \sqrt{3}}\epsilon + \dfrac{16}{243 \sqrt{3}} \epsilon^2$} \\ 
                & Rezzolla-Zhidenko & $-\dfrac{16}{9}\epsilon + \dfrac{28892}{9801} \epsilon^2$ & \raisebox{0.015cm}{$\dfrac{8}{81\sqrt{3}}\epsilon - \dfrac{4012}{24057\sqrt{3}} \epsilon^2$}& \raisebox{0.015cm}{$-\dfrac{16}{81\sqrt{3}} \epsilon + \dfrac{173960}{2910897 \sqrt{3}} \epsilon^2 $}\\ 
                & Johannsen-Psaltis &$-\dfrac{16}{9} \epsilon+ \dfrac{32}{9} \epsilon^2$ & \raisebox{0.015cm}{$\dfrac{8}{81 \sqrt{3}} \epsilon -\dfrac{440 }{2187 \sqrt{3}} \epsilon ^2$}& \raisebox{0.015cm}{$ -\dfrac{16 }{81 \sqrt{3}} \epsilon + \dfrac{32 }{243 \sqrt{3}} \epsilon ^2 $} \\
                & EMD near horizon & $-\dfrac{16}{9} \epsilon + \dfrac{799}{243} \epsilon^2$ & \raisebox{0.015cm}{$\dfrac{8}{81 \sqrt{3}} \epsilon- \dfrac{110}{729\sqrt{3}} \epsilon^2$} & \raisebox{0.015cm}{$-\dfrac{16}{81\sqrt{3}} \epsilon -\dfrac{107}{2187\sqrt{3}} \epsilon^2 $}\\ 
                & EMD large distance & $-\dfrac{16}{9} \epsilon + \dfrac{176 }{81} \epsilon ^2$ & \raisebox{0.015cm}{$\dfrac{8}{81 \sqrt{3}} \epsilon -\dfrac{296 }{2187 \sqrt{3}} \epsilon ^2$}& \raisebox{0.015cm}{$-\dfrac{16}{81\sqrt{3}} \epsilon -\dfrac{368 }{2187 \sqrt{3}} \epsilon ^2$} \\[0.2cm]
        \hline
        \multirow{5}{*}{\raisebox{-0.2cm}{Bardeen}} 
        & From the model & $-\dfrac{20}{9}\epsilon + \dfrac{650}{243}\epsilon^2$ 
                & $\dfrac{4}{27 \sqrt{3}}\epsilon - \dfrac{98}{729 \sqrt{3}}\epsilon^2 $ 
                & $ -\dfrac{8}{81 \sqrt{3}}\epsilon - \dfrac{68}{729 \sqrt{3}} \epsilon^2 $ \\ 
                & Rezzolla-Zhidenko & $-\dfrac{20}{9}\epsilon + \dfrac{650}{243}\epsilon^2$ & $\dfrac{4}{27 \sqrt{3}}\epsilon - \dfrac{98}{729 \sqrt{3}}\epsilon^2 $ & $ -\dfrac{8}{81 \sqrt{3}}\epsilon - \dfrac{68}{729 \sqrt{3}} \epsilon^2 $\\ 
                & Johannsen-Psaltis & $-\dfrac{20}{9}\epsilon + \dfrac{650}{243}\epsilon^2$ & $\dfrac{4}{27 \sqrt{3}}\epsilon - \dfrac{98}{729 \sqrt{3}}\epsilon^2 $ & $ -\dfrac{8}{81 \sqrt{3}}\epsilon - \dfrac{68}{729 \sqrt{3}} \epsilon^2 $ \\
                & EMD near horizon & $-\dfrac{20}{9}\epsilon+\dfrac{652 }{243}\epsilon ^2$ & $\dfrac{4}{27 \sqrt{3}} \epsilon- \dfrac{391}{2916\sqrt{3}} \epsilon^2$ & $-\dfrac{8}{81\sqrt{3}} \epsilon -\dfrac{421}{4374\sqrt{3}} \epsilon^2 $ \\ 
                & EMD large distance & $-\dfrac{20}{9} \epsilon + \dfrac{370 }{243} \epsilon ^2$ & $\dfrac{4}{27 \sqrt{3}} \epsilon -\dfrac{98 }{729 \sqrt{3}} \epsilon ^2$& $-\dfrac{8}{81\sqrt{3}} \epsilon - \dfrac{68}{729 \sqrt{3}} \epsilon ^2$ \\[0.2cm]
        \hline
        \multirow{1}{*}{\raisebox{-0.05cm}{Simpson-Visser II}} 
        &   \multirow{1}{*}{\raisebox{-0.05cm}{All parametrizations}} & \raisebox{0.05cm}{$-\dfrac{8}{3}\epsilon + \dfrac{74}{27}\epsilon^2 $} 
                & \raisebox{0.07cm}{$\dfrac{2}{9 \sqrt{3}}\epsilon - \dfrac{7}{81 \sqrt{3}}\epsilon^2 $} 
                & \raisebox{0.07cm}{$\dfrac{2}{27 \sqrt{3}}\epsilon - \dfrac{11}{81\sqrt{3}} \epsilon^2$} \\[0.1cm]
        \hline
    \end{tabular}
      \caption{\justifying Expressions of the photon sphere radius $\rps$, orbital frequency $\Omega$, and Lyapunov exponent $\lambda$. The values are rescaled by the BH mass $M$, shifted with respect to the corresponding Schwarzschild result, and computed up to quadratic order in $\epsilon$ in the case of Hayward, Bardeen, and Simpson-Visser II models using four parameters in each parametrization. For the last BH spacetime, we report only the expressions for the exact metric, as they are identical across all parametrizations.}
      \label{tab:BH-models}
\end{table*}

We analytically expand the quantities in Eqs.~\eqref{eq:EIKONAL-QUANTITIES} around $\epsilon = 0$ up to quadratic order for each BH model and compare the results with those obtained in the Rezzolla–Zhidenko, Johannsen–Psaltis, near-horizon EMD, and large-distance EMD parametrizations, fixing the number of free parameters to four in Eqs.~\eqref{eikonal q rz}, \eqref{eikonal q jp}, \eqref{eikonal q emd nh}, and \eqref{eikonal q emd ld}. The outcome of this comparison is summarized in Table~\ref{tab:BH-models}.
We observe that, at linear order, all parametrizations reproduce the same result for all physical quantities in all BH models. Aside from the Simpson–Visser~II case, where all parametrizations trivially yield identical expansions, no single parametrization consistently performs better than the others across all observables and BH models.
\begin{table*}[ht]
    \centering
    \renewcommand{\arraystretch}{2} % increase vertical spacing between rows
\begin{tabular}{|wc{3cm}|wc{1cm}|wc{1cm} |wc{1cm} |wc{1cm} |wc{1.4cm} |wc{1.4cm}| }
\hline
\raisebox{0.05cm}{BH models}  & \multicolumn{2}{c|}{\raisebox{0.07cm}{Hayward}} & \multicolumn{2}{c|}{\raisebox{0.07cm}{Bardeen}} & \multicolumn{2}{c|}{\raisebox{0.07cm}{Simpson-Visser II}} \\ 
\hline
\raisebox{0.07cm}{order} &\raisebox{0.07cm}{$\epsilon$} &\raisebox{0.07cm}{$\epsilon^2$} &  \raisebox{0.07cm}{$\epsilon$}         &      \raisebox{0.07cm}{$\epsilon^2$}         &       \raisebox{0.07cm}{$\epsilon$}    &        \raisebox{0.07cm}{$\epsilon^2$}      \\
\hline
Rezzolla-Zhidenko  &   2        &      7     &     2      &      4     &    1       &   1        \\
Johannsen-Psaltis  &     3      &   6        &      2  &      4     &     1      &       2    \\
EMD near horizon &     4      &      $>7$     &         4  &       $>7$     &   4        & 4        \\
EMD large distance &    3      &     6    &     2      &     4      &     1      &       2  \\[0.12cm]
\hline
\end{tabular}
    \caption{\justifying Number of parameters required to reproduce the exact results for the three observables at orders $\epsilon$ and $\epsilon^2$.}
     \label{tab:NumbParam}
\end{table*}    

\begin{figure*}[ht]
    \centering
    \includegraphics[width=\textwidth]{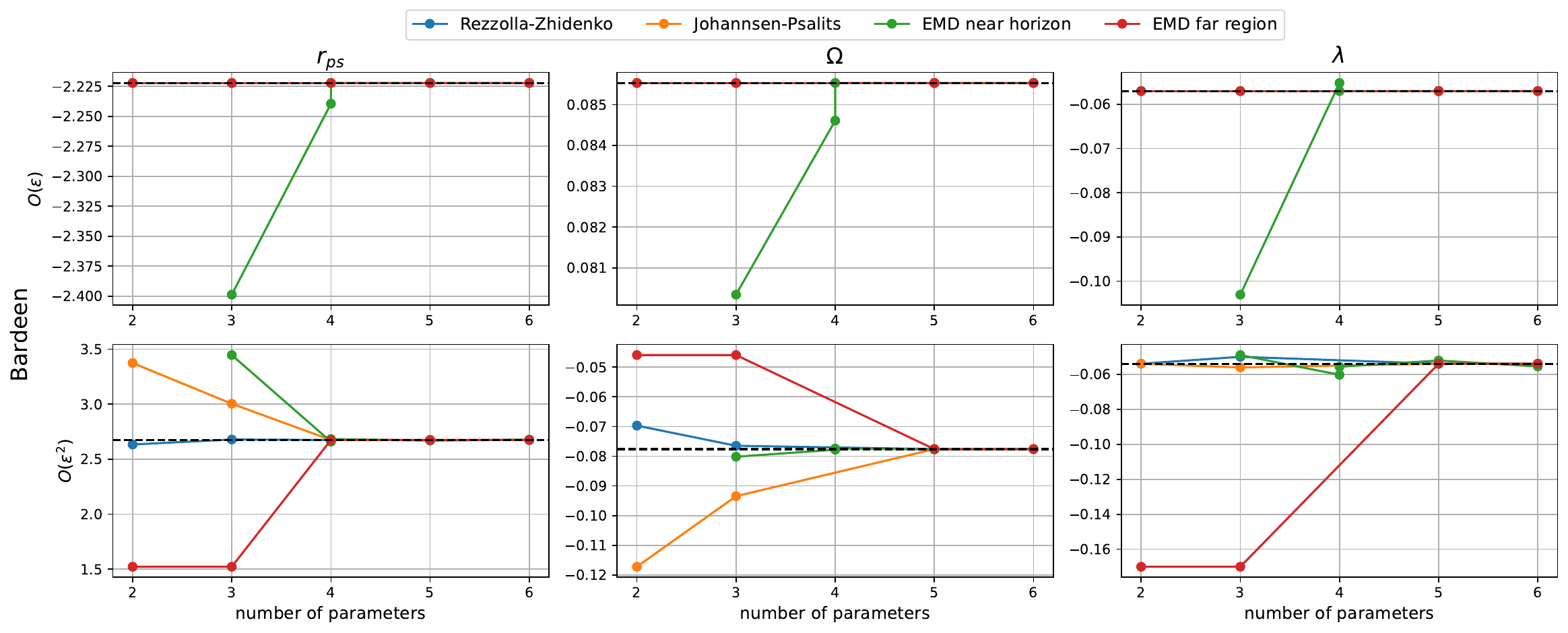}
    \caption{\justifying Comparison of the convergence to the linear coefficient (top row) and quadratic coefficient (bottom row) for the Bardeen BH model, varying the number of the related parameters. The horizontal dashed black line is the value computed from the model, reported in Table~\ref{tab:BH-models}. For the near-horizon EMD with four parameters, two Padé approximants were used: (2,2) and (1,3), the latter being more accurate.}
    \label{fig:bardeen_conv}
\end{figure*}

A complementary analysis is presented in Table~\ref{tab:NumbParam}, where, instead of fixing the number of parameters a priori, we set the desired tolerance and let the required number of parameters vary to match the exact expressions. In the case of the EMD near-horizon parametrization, this procedure coincides with a Padé approximation of order $(N,3)$, where the number of free parameters is controlled by the integer $N$. Also in this setting, no parametrization emerges as universally favored across all BH models.

Finally, in Fig.~\ref{fig:bardeen_conv} we illustrate the convergence of the linear and quadratic coefficients for the Bardeen BH as a representative example. Similar behavior is observed for all other BH models. The plots highlight that all parametrizations perform on a comparable footing. Therefore, the choice of parametrization should be guided by the desired accuracy and the number of parameters one is willing to introduce. This has to be considered on a case-by-case basis for each physical observable and BH model.

\section{Discussion and conclusions}
\label{sec:end}
This work begins by examining three model-independent parametrizations of BH spacetimes: Johannsen–Psaltis, which describes deformations as quadrupolar deviations from GR; Rezzolla–Zhidenko, which introduces departures from the Schwarzschild solution and employs Padé-like continued-fraction expansions valid far from the event horizon; and EMD, which provides local BH metric descriptions in terms of physical quantities such as the physical distance. We then establish, for the first time, explicit mappings among these three frameworks, thereby offering a unified perspective on agnostic parametrizations of deformed BHs. 

As a relevant application, we investigate the QNMs of BHs in the eikonal limit. We analyse the QNM spectrum across the different parametrizations, considering small deviations from the Schwarzschild case by treating the deformation parameter  ($\epsilon$) as a perturbative quantity. This allows us to expand the effective coefficients of each parametrization in power series of $\epsilon$. To demonstrate the effectiveness of our framework, we apply it to three regular BH models: Hayward, Bardeen, and Simpson–Visser II. We summarize our results in several tables and figures carrying different information: in Table \ref{tab:BH-models} we report the expansions up to quadratic order in $\epsilon$ using only four parameters for each parametrization and for each BH model; in Table \ref{tab:NumbParam} we fix the accuracy and let the number of coefficients vary freely; in Fig. \ref{fig:bardeen_conv} we show the convergence of the different expansions in terms of the number of retained parameters and also check the values of the attained tolerances. Across all analyses, we find that the three parametrizations are fundamentally equivalent: none systematically outperforms the others. Superiority is context-dependent and emerges only once the target accuracy, the observable under consideration, and the number of parameters are specified within a given BH model.

Our astrophysical case study supports a democratic perspective: parametrizations should be viewed as complementary tools rather than competing prescriptions. This philosophy naturally paves the way for developing new parametrizations tailored to specific observables or regimes. Moreover, recent works~\cite{DamiaPaciarini:2025xjc, Hohenegger:2025dic} have emphasized the usefulness of proper-time EMD parametrizations in capturing effective quantum corrections near the horizon. In this light, the mappings established in our work offer promising opportunities to translate such quantum-gravity–inspired deformations across different parametrization schemes. As a natural extension, the same methodology could be applied to other problems in BH physics and generalized to stationary and axially symmetric spacetimes.  

\section*{Acknowledgements}
V. De Falco is grateful to Gruppo Nazionale di Fisica Matematica of Istituto Nazionale di Alta Matematica for support. C. De Simone acknowledges the support of INFN sez. di Napoli, \emph{iniziativa specifica} QGSKY. M. Del Piano acknowledges support from INFN sez. di Napoli, \emph{iniziativa specifica} ST\&FI and expresses gratitude to the Southern Denmark University, the Danish Institute for Advanced Study, and the Quantum Theory Center in Odense (Denmark) for hosting during the initial stages of the work. The work of F. Sannino and M. Damia Paciarini is partially supported by the Carlsberg Foundation, grant CF22-0922. This work contributes to COST Action CA23130 -- Bridging high and low energies in search of quantum gravity (BridgeQG).

\appendix
\onecolumngrid
\section{Treatment of logarithmic functions for EMD at large distance} \label{app:EMDlogs}

\renewcommand{\arraystretch}{3.5}
\setlength{\tabcolsep}{10pt}
\begin{table*}[ht]
    \centering

\scalebox{0.9}{
    \begin{tabular}{|c|c|c|c|c|}
        \hline
        \raisebox{0.5cm}{$N$} &  \raisebox{0.5cm}{$M$} &  \raisebox{0.51cm}{$\eta_0$} & \multicolumn{1}{c|}{ \raisebox{0.51cm}{$\eta_1$}} \\[-0.7cm]
        \hline
        \raisebox{0.2cm}{2} &\raisebox{0.2cm}{1} & \raisebox{0.22cm}{$ \dfrac{1}{4} \left(\sqrt{3}-1\right)$} 
          & \raisebox{0.21cm}{$-\dfrac{1}{4} \left(\sqrt{3}-1\right) 
          +\dfrac{1}{135} \left(61-68 \sqrt{3}\right) \dfrac{\theta_2}{\mathfrak{c}} 
          +\dfrac{2}{9} \left(2 \sqrt{3}-7\right) \dfrac{\theta_4}{\mathfrak{c}} 
          +\dfrac{8}{3} \left(\sqrt{3}-2\right) \dfrac{\theta_6}{\mathfrak{c}}$} \\[-0.1cm]
        \hline
      \multirow{2}{0.01\textwidth}{\raisebox{0.06cm}{2}} & \multirow{2}{0.01\textwidth}{ \raisebox{0.45cm}{2}} & \multirow{2}{0.2\textwidth}{\centering{\raisebox{0.27cm}{$\dfrac{1}{4} \sqrt{9 \sqrt{3}-15}$}} }& \multirow{2}{0.69\textwidth}{\hspace{0.6cm}\raisebox{0.3cm}{$-\dfrac{1}{4} \sqrt{9 \sqrt{3}-15}-\dfrac{1}{630} \sqrt{1274659 \sqrt{3}-\dfrac{6386663}{3}}\dfrac{\theta_2}{\mathfrak{c}}-\dfrac{11}{15} \sqrt{\dfrac{1}{6} \left(627 \sqrt{3}-1075\right)}\dfrac{\theta_4}{\mathfrak{c}}$}\\

      \hspace{1.9cm}\raisebox{0.7cm}{$- \dfrac{2444\sqrt{2}}{\sqrt{\left(14 \sqrt{3}-9\right) \left(85675 \sqrt{3}+148401\right)}}\dfrac{\theta_6}{\mathfrak{c}}-\dfrac{16}{\sqrt{51 \sqrt{3}+\dfrac{265}{3}}}\dfrac{\theta_8}{\mathfrak{c}}$}} \\[0.1cm]
   & & &\\[-0.2cm]
        \hline
        \multirow{1}{0.01\textwidth} {1} & \multirow{1}{0.01\textwidth} {3} & \multirow{1}{0.2\textwidth} {\hspace{0.06cm}\centering{$\dfrac{1}{3 \sqrt{3}}$} }
          & \multirow{1}{0.69\textwidth} {\centering{ $-\dfrac{1}{3 \sqrt{3}} 
            -\dfrac{19676}{25515 \sqrt{3}}\dfrac{\theta_2}{\mathfrak{c}} 
            -\dfrac{656}{405 \sqrt{3}}\dfrac{\theta_4}{\mathfrak{c}} 
            -\dfrac{64}{27 \sqrt{3}}\dfrac{\theta_6}{\mathfrak{c}} 
        -\dfrac{128}{81 \sqrt{3}}\dfrac{\theta_8}{\mathfrak{c}}$}} \\
        \hline
       \multirow{1}{0.01\textwidth} {2} & \multirow{1}{0.01\textwidth} {3} & \multirow{1}{0.2 \textwidth}{ \centering{\hspace{-0.16cm}$\dfrac{1}{3 \sqrt{3}}$}} 
          & \multirow{1}{0.69 \textwidth}{\centering{$-\dfrac{1}{3 \sqrt{3}} 
            -\dfrac{98837}{127575 \sqrt{3}}\dfrac{\theta_2}{\mathfrak{c}} 
            -\dfrac{136096}{76545 \sqrt{3}}\dfrac{\theta_4}{\mathfrak{c}} 
            -\dfrac{4384}{1215 \sqrt{3}}\dfrac{\theta_6}{\mathfrak{c}} 
            -\dfrac{1216}{243 \sqrt{3}}\dfrac{\theta_8}{\mathfrak{c}} 
            -\dfrac{256}{81 \sqrt{3}}\dfrac{\theta_{10}}{\mathfrak{c}}$}} \\
        \hline
         \multirow{2}{0.01\textwidth}{ \raisebox{0.6cm}{3}} & \multirow{2}{0.01\textwidth}{ \raisebox{0.6cm}{3}} & \multirow{2}{0.2\textwidth}{\hspace{-0.05cm}\centering{\raisebox{0.47cm}{$\dfrac{1}{3 \sqrt{3}}$}}} 
          & \raisebox{0.23cm}{$-\dfrac{1}{3 \sqrt{3}}-\dfrac{652294}{841995 \sqrt{3}}\dfrac{\theta_2}{\mathfrak{c}}-\dfrac{2066276}{1148175 \sqrt{3}}\dfrac{\theta_4}{\mathfrak{c}}-\dfrac{62528}{15309 \sqrt{3}}\dfrac{\theta_6}{\mathfrak{c}}-\dfrac{29152}{3645 \sqrt{3}}\dfrac{\theta_8}{\mathfrak{c}}$}\\
            & & &\raisebox{0.5cm}{$-\dfrac{2560}{{243 \sqrt{3}}}\dfrac{\theta_{10}}{\mathfrak{c}}-\dfrac{512}{81 \sqrt{3}}\dfrac{\theta_{12}}{\mathfrak{c}}$}\\[-0.35cm]
        \hline
        \hline        
        \raisebox{0.5cm}{$N$} & \raisebox{0.5cm}{$M$} & \raisebox{0.5cm}{$\Upsilon_0$} & \multicolumn{1}{c|}{\raisebox{0.5cm}{$\Upsilon_1$}} \\[-0.7cm]
        \hline
         \multirow{2}{0.01\textwidth}{ \raisebox{0.5cm}{2}} & \multirow{2}{0.01\textwidth}{ \raisebox{0.5cm}{1}} &\multirow{2}{0.2\textwidth}{\hspace{0.24cm}\raisebox{0.49cm}{ $\dfrac{\left(\sqrt{3}-1\right) \left(\sqrt{3}+1\right)^2}{16 \sqrt[4]{3}}$} 
          }& 
          
          \raisebox{0.1cm}{ $-\dfrac{\left(\sqrt{3}-1\right) \left(\sqrt{3}+1\right)^2}{16 \sqrt[4]{3}}+\dfrac{\left(13 \sqrt{3}-116\right) \left(\sqrt{3}+1\right)}{180 \sqrt[4]{3}}\dfrac{\theta_2}{\mathfrak{c}}+\dfrac{\left(\sqrt{3}-2\right) \left(\sqrt{3}+1\right)}{2 \sqrt[4]{3}}\dfrac{\theta_4}{\mathfrak{c}}$} \\
          & & & \raisebox{0.3cm}{ $-\dfrac{2 \left(\sqrt{3}-2\right) \left(\sqrt{3}+1\right)}{\sqrt[4]{3}}\dfrac{\theta_6}{\mathfrak{c}}$}\\[-0.2cm]
        \hline
         \multirow{2}{0.01\textwidth}{ \raisebox{0.45cm}{2}} & \multirow{2}{0.01\textwidth}{ \raisebox{0.45cm}{2}} &\multirow{2}{0.2\textwidth}{ \hspace{0.1cm}\raisebox{0.38cm}{$\dfrac{1}{24} \sqrt{\dfrac{1}{2} \left(963-529 \sqrt{3}\right)}$}}& \raisebox{0.15cm}{$
            -\dfrac{1}{24} \sqrt{\dfrac{1}{2} \left(963-529 \sqrt{3}\right)}
            -\dfrac{\sqrt{187519473-102837005 \sqrt{3}}}{3780}\dfrac{\theta_2}{\mathfrak{c}}
            +\dfrac{1}{60} \sqrt{162387-92399 \sqrt{3}}\dfrac{\theta_4}{\mathfrak{c}}$}\\
            & & &
             \raisebox{0.45cm}{+$\dfrac{78 \left(9 \sqrt{3}+23\right)}{\sqrt{61617 \sqrt{3}+106731}}\dfrac{\theta_6}{\mathfrak{c}}+4796 \sqrt{\dfrac{2}{241053 \sqrt{3}+417609}}\dfrac{\theta_8}{\mathfrak{c}}$ }\\[-1.5cm]
   & & &\\
        \hline
        \raisebox{0.16cm}{1} & \raisebox{0.16cm}{3} & \raisebox{0.2cm}{$\dfrac{1}{3 \sqrt{3}}$} 
          & \raisebox{0.19cm}{$-\dfrac{1}{3 \sqrt{3}} 
            -\dfrac{35396}{25515 \sqrt{3}}\dfrac{\theta_2}{\mathfrak{c}} 
            +\dfrac{112}{1215 \sqrt{3}}\dfrac{\theta_4}{\mathfrak{c}} 
            +\dfrac{448}{81 \sqrt{3}}\dfrac{\theta_6}{\mathfrak{c}} 
            +\dfrac{512}{81 \sqrt{3}}\dfrac{\theta_8}{\mathfrak{c}}$} \\
        \hline
       \multirow{1}{0.01\textwidth}{\raisebox{0.0cm}{2}}  & \multirow{1}{0.01\textwidth}{\raisebox{0.0cm}{3}} & \multirow{1}{0.2\textwidth}{\centering{\hspace{0.1cm}$\dfrac{1}{3 \sqrt{3}}$} 
         } &\multirow{1}{0.69\textwidth}{ \centering{$-\dfrac{1}{3 \sqrt{3}} 
            -\dfrac{171496}{127575 \sqrt{3}}\dfrac{\theta_2}{\mathfrak{c}} 
            +\dfrac{10160}{5103 \sqrt{3}}\dfrac{\theta_4}{\mathfrak{c}} 
            +\dfrac{2752}{135 \sqrt{3}}\dfrac{\theta_6}{\mathfrak{c}} 
            +\dfrac{1280}{27 \sqrt{3}}\dfrac{\theta_8}{\mathfrak{c}} 
            +\dfrac{1024}{27 \sqrt{3}}\dfrac{\theta_{10}}{\mathfrak{c}}$}} \\
        \hline
         \multirow{2}{0.01\textwidth}{ \raisebox{0.35cm}{3}} & \multirow{2}{0.01\textwidth}{ \raisebox{0.35cm}{1}} &\multirow{2}{0.2\textwidth}{\centering{ \raisebox{0.1cm}{$\dfrac{1}{3 \sqrt{3}}$} }}
          & \raisebox{0.12cm}{$-\dfrac{1}{3 \sqrt{3}} 
            -\dfrac{5662841}{4209975 \sqrt{3}}\dfrac{\theta_2}{\mathfrak{c}} 
            +\dfrac{2857228}{1148175 \sqrt{3}}\dfrac{\theta_4}{\mathfrak{c}} 
            +\dfrac{2398688}{76545 \sqrt{3}}\dfrac{\theta_6}{\mathfrak{c}} 
            +\dfrac{423776}{3645 \sqrt{3}}\dfrac{\theta_8}{\mathfrak{c}}$} \\
            & & & \raisebox{0.32cm}{$+\dfrac{50432}{243 \sqrt{3}}\dfrac{\theta_{10}}{\mathfrak{c}} 
            +\dfrac{11776}{81 \sqrt{3}}\dfrac{\theta_{12}}{\mathfrak{c}}$} \\[-0.1cm]
        \hline
        \hline
    \end{tabular}
    }
    
    \caption{ \justifying Orbital frequency $(\eta_0,\eta_1)$ and Lyapunov exponent $(\Upsilon_0,\Upsilon_1)$ coefficients of the photon sphere obtained from Eqs. \eqref{eq:orbital_frequency} and \eqref{eq:Lyapunov_exponent} in terms of the Padé approximants of the potential $U^2_{N,M}$.} 
    \label{Tab:potential_Padé}
\end{table*}
In order to determine the metric functions, we could start by assuming that $\Phi(d)$ and $\Psi(d)$ can be expanded in power series of $1/d$. However, such an approach will lead to logarithmic terms in the asymptotic expansions of $h(r)$ and $f(r)$. Therefore, to avoid that consequence, we can reabsorb the logarithms in $\Phi(d)$ and $\Psi(d)$ by defining them as
\begin{align}\label{phipsilogs}
        \Phi(d) &= 2M \left[ 1 + \sum_{n=1}^\infty \sum_{m=0}^{n-1} \frac{\newom_{n,m} M^n}{d^n} \log^{m}\left(\frac{d}{2M} \right)\right] \, , \quad
        \Psi(d) = 2M \left[ 1 + \sum_{n=1}^\infty \sum_{m=0}^{n-1} \frac{\newgam_{n,m} M^n}{d^n} \log^{m}\left(\frac{d}{2M} \right)\right]\ ,
\end{align}
where $\newom_{n,m}$ and $\newgam_{n,m}$ are real constants and $\omega_n \equiv \newom_{n,0}$ and $\gamma_n \equiv \newgam_{n,0}$.

We consider the post-Newtonian (PN) expansion  $M/r\ll1$, (the weak-field limit $M/r\ll1$ is related to low velocity limit $v/c\ll1$ via the virial theorem \cite{Maggiore:2007ulw}, namely $M/r\sim v^2/c^2$). In this regime, we have $1/(r^{i}d^j)\sim 1/(r^{i+j})$ and the function $d(r)$ admits the logarithmic corrections: $(M/r)^{j}\log^k(r/M)$, with $\{k,j\} \in \mathbb{N}_0$ and $0 < k \leq j-1$. We PN-expand Eq. \eqref{eqprop} via Eq. \eqref{phipsilogs}, which gives \cite{DelPiano:2024gvw, DAlise:2023hls}
\begin{align} \label{largedistexplogs}
    \frac{d(r)}{M}&=\frac{r}{M}+\log\left(\frac{r}{2M} \right) + k -\left(\frac{3}{2}+\omega_1\right)\frac{M}{r} 
    - \left[ \frac{5}{4}+\frac{\omega_1(5-2k)}{4} + \frac{\omega_2+ \newom_{2,1}}{2} \right] \frac{M^2}{r^2} +\frac{M^2(\omega_1 - \newom_{2,1}) }{2r^2}\log\left(\frac{r}{2M} \right) \notag \\
    &\!\!\!\!\!\!\!\!- \left[\frac{35}{24} + \frac{(74 - 21 k + 9 k^2)\omega_1}{27} + \frac{45 \omega_1^2 + 6 (7-6k) \omega_2 + 2(8+3k)\newom_{2,1}}{54 } + \frac{9\omega_3+3 \newom_{3,1} + 2\newom_{3,2}}{27} \right] \frac{M^3}{r^3}+\Bigg{[}(7-6k) \omega_1 \notag\\
    &+ 6 \omega_2-2(4-3k)\newom_{2,1}- 3\newom_{3,1}-2\newom_{3,2}\Bigg{]} \frac{M^3}{9r^3} \log\left( \frac{r}{2M}\right)- [ \omega_1 + 2 \newom_{2,1}-\newom_{3,2}]\frac{M^3}{3r^3} \log ^2\left( \frac{r}{2M}\right)+\order{\frac{M^4}{r^4}} \ ,
    \end{align}
where $k$ is a real integration constant, kept as a free parameter. It is important to note that $k$ plays the role of the gauge for $d$, as its value determines the position at which $d=0$. Inserting Eq. \eqref{largedistexplogs} into the metric functions modified with the series in Eq.~\eqref{phipsilogs}, we obtain the conditions to balance and cancel out (up to a given order) the logarithmic functions in Eq.~\eqref{eq:EMD-functions}, which requires
\begin{align}
    \newom_{2,1} &= \omega_1 \ , \ \newom_{3,1} = -  \omega_1 + 2 \omega_2 \ ,   \ \newom_{3,2} =  \omega_1 \ , \ %\\
     \newom_{4,3}=\omega_1 \ ,\  \newom_{4,2}=-\frac{5}{2}  \omega_1 + \frac{6}{2} \omega_2 \ ,\ 
     \newom_{4,1} =\omega_1 -2  \omega_2 + 3  \omega_3 \ ,  \notag \\
     \newom_{5,4} &= \omega_1 \ , \ \newom_{5,3} = -\frac{13}{3}  \omega_1 + \frac{12}{3} \omega_2 \ ,  \ \newom_{5,2} = \frac{9}{2} \omega_1 - \frac{14}{2} \omega_2 + \frac{12}{2} \omega_3 \ ,\ \newom_{5,1} = - \omega_1 + 2  \omega_2 - 3  \omega_3 + 4  \omega_4 \ , \notag
\end{align}
for the first few terms. Although it is difficult to generalize the above relations, they can be computed to arbitrary order and applied to the set $\{ \newgam_{n,m} \}_{n\geq 1}^{0 \leq m \leq n-1}$. Using the above conditions, we are able to obtain the metric functions as in Eq. \eqref{eq:EMD-functions} and eliminate the logarithmic functions appearing in Eq.~\eqref{largedistexplogs}.

\section{Extending the EMD near horizon in the eikonal limit}\label{app:emd eikonal}
In the EMD framework near the horizon, the potential $U^2(r)$ can be expressed as a power series around $\hr$:
    \begin{equation}\label{u_series}
        U^2(r)=\sum_{n=1}^{\infty}v_n (r-\hr)^n,
    \end{equation}
where $v_n$ depend on $\{\theta_{2k}\}$ and $\{\xi_{2k}\}$ (see \cite{DelPiano:2024nrl} for the explicit form of these coefficients). It has been shown in~\cite{DelPiano:2024nrl} that, by employing Padé approximations at a sufficiently high order $(N,M)$, the convergence radius of the series increases significantly up to the photon sphere location. This approach allows for the computation of the BH shadow radius $\bsh$, orbital frequency $\Omega$, and Lyapunov exponent $\lambda$. Assuming that the position of the BH horizon is defined by Eq.~\eqref{r_horizonEMD}, for small deviations $\mathfrak{c} \ll 1$ from the Schwarzschild horizon, we can write $\Omega$ and $\lambda$ in mass units $M$ as:
    \begin{subequations}\label{eq:omegal}
    \begin{align}
        \Omega M&=\eta_0+\eta_1\;\mathfrak{c}+\mathcal{O}(\mathfrak{c}^2),\\
        \lambda\, M&=\Upsilon_0+\Upsilon_1\;\mathfrak{c}+\mathcal{O}(\mathfrak{c}^2).
    \end{align}        
    \end{subequations}
The coefficients $(\eta_0,\eta_1)$ and $(\Upsilon_0,\Upsilon_1)$ are reported in Table \ref{Tab:potential_Padé} and depend on the order of the Padé approximation $(N,M)$. The Schwarzschild limit is already recovered at order $(1,3)$. From Eq. \eqref{eq:omegal}, the eikonal QNMs can be directly reconstructed at any order $(N,M)$ of the Padé approximation using Eq. \eqref{eq:eikonal}.

\begin{figure*}[ht]
        \centering
        \hbox{\includegraphics[width=\linewidth]{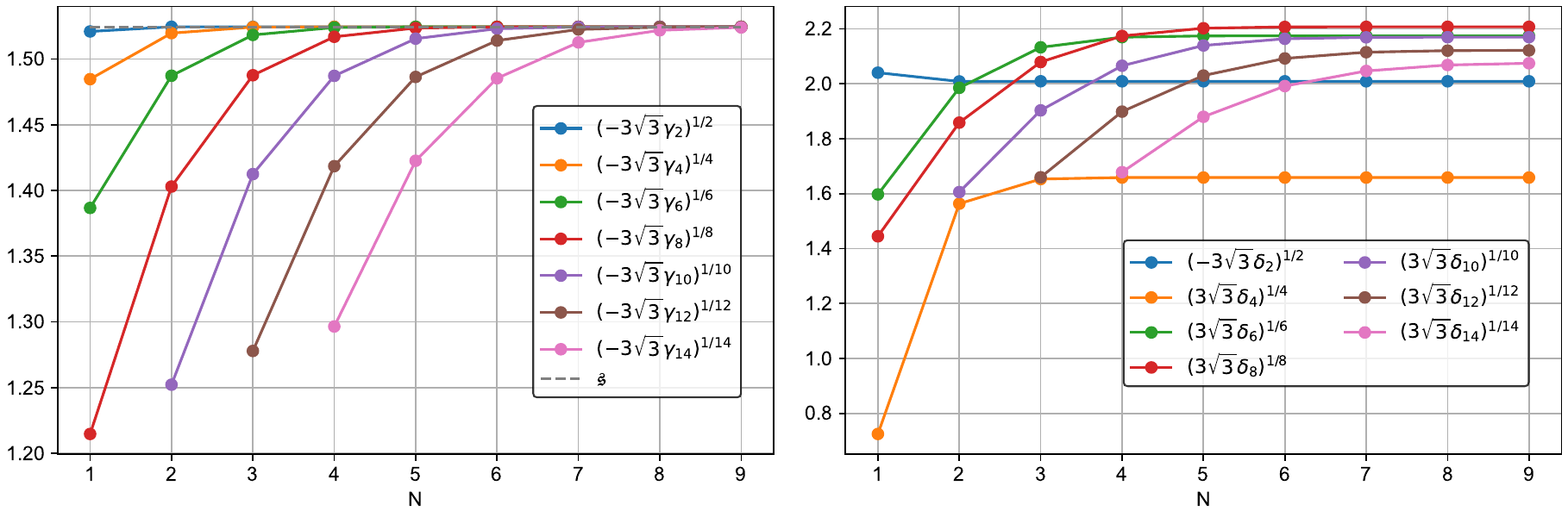}} 
        \caption{ \justifying Coefficients $\gamma_{2k}$ (left) and $\delta_{2k}$ (right) for different orders of the Padé approximation $(N,3)$.}
        \label{fig:QNM_frequency}
\end{figure*}

The relation between $\bsh$ and $\Omega$ can be further exploited to investigate the large-$N$ limit in the Padé approximation of order $(N,3)$. First, let us recall the expression of the shadow radius in terms of $\mathfrak{c}$ from \cite{DelPiano:2024nrl}:
    \begin{equation}
          \frac{b_{\rm{sh}}}{M}=\tau_0+\tau_1\;\mathfrak{c}+\mathcal{O}(\mathfrak{c}^2),
    \end{equation}
as well as the explicit dependence of $\tau_1$ on $\mathfrak{c}$ and on the order $N$ of the Padé approximation:
    \begin{align}
        \tau_1(N)&=3\sqrt{3}+\sum_{k=1}^{N+3} \beta_{2k}(N)\frac{\theta_{2k}}{\mathfrak{c}} \;\;\;\; \qq{with}   \;\;\;\;      \beta_{2k}(N)=\frac{\partial}{\partial \,\theta_{2k}}\frac{1}{U_{N,3}}\bigg|_{(\rps)_{N,3}}, 
    \end{align}
where $(\rps)_{N,3}$ and $U_{N,3}$ correspond to the photon sphere radius and potential computed at the order $(N,3)$ of the Padé approximation, respectively. Similarly, the coefficient $\eta_1$ of the photon sphere frequency can be expressed as
    \begin{align}
        \eta_1(N)&=-\frac{1}{3\sqrt{3}}+\sum_{k=1}^{N+3} \gamma_{2k}(N)\frac{\theta_{2k}}{\mathfrak{c}}
        \;\;\;\; \qq{with}   \;\;\;\;         \gamma_{2k}(N)=\frac{\partial \,U_{N,3}}{\partial \,\theta_{2k}}\bigg|_{(\rps)_{N,3}} .
    \end{align}
Recalling that, at the lowest order, $U_{N,3}=1/(3\sqrt{3})$, the coefficients $\gamma_{2k}(N)$ and $\beta_{2k}(N)$ are related by 
\begin{align}
        \beta_{2k}(N)&=\frac{\partial}{\partial \,\theta_{2k}}\frac{1}{U_{N,3}}\bigg|_{(r_{\rm ps})_{N,3}}=-\frac{1}{U^2_{N,3}}\gamma_{2k}(N)=-(3\sqrt{3})^2\gamma_{2k}(N),
    \end{align}
and the asymptotic behavior of $\gamma_{2k}(N)$ can be derived from the $N\to \infty$ limit of $\beta_{2k}(N)$ obtained in \cite{DelPiano:2024nrl}
    \begin{equation}
        \lim_{N\to\infty} \frac{\beta_{2k}(N)}{3\sqrt{3}}\to\mathfrak{s}^{2k}, \;\;\;\;\;\;\;\mathfrak{s}\sim 1.5245.
    \end{equation}
This result is compatible with Fig. \ref{fig:QNM_frequency}, where the first few coefficients $\gamma_{2k}$ have been computed up to $N=9$.
Moreover, the asymptotic limit of $\eta_1(N)$ becomes
    \begin{align}
        \lim_{N\to\infty} \mathfrak{c}\,\eta_1(N)&=-\frac{\mathfrak{c}}{3\sqrt{3}}-\sum_{k=1}^\infty\frac{\mathfrak{s}^{2k}}{3\sqrt{3}}\theta_{2k}
        =-\frac{\mathfrak{c}}{3\sqrt{3}}-\frac{1}{3\sqrt{3}}\frac{1}{2M}[\Psi(2M\mathfrak{s})-\hr],
    \end{align}
and substituting it into the full expression of $\Omega=\eta_0+\eta_1\mathfrak{c}$, we obtain
    \begin{align}
       \lim_{N\to\infty} \Omega&=\lim_{N\to\infty}(\eta_0+\eta_1 \mathfrak{c})=
        \frac{2}{3\sqrt{3}}-\frac{1}{3\sqrt{3}}\frac{1}{2M}\Psi(2M\mathfrak{s}).
    \end{align}
In the case of the Lyapunov exponent, the asymptotic $N$ limit is more intricate, since $\lambda$ depends on both the photon sphere radius and the second-order derivatives of the potential. However, by analogy with $\Omega$, we adopt a similar ansatz for $\Upsilon_1(N)$, which can be written as
    \begin{equation}
        \Upsilon_1(N)=\frac{1}{3\sqrt{3}}+\sum_{k=1}^{N+3}\delta_{2k}(N)\frac{\theta_{2k}}{\mathfrak{c}},
    \end{equation}
where $\delta_{2k}$ is a suitable set of numerical coefficients. In general, $\delta_{2k}$ depend on the order of the Padé approximation and are related to the derivative of the Lyapunov exponent with respect to the parameters $\theta_{2k}$, evaluated at the radius of the photon sphere. The coefficients $\delta_{2k}$ computed up to $k=7$ and $N=9$ are shown in Fig. \ref{fig:QNM_frequency}. Although the individual values of $\delta_{2k}$ seem to converge to a constant value, it is not clear whether resummation similar to the BH shadow radius and photon sphere frequency is possible. We note that the coefficient $\delta_8$ reaches the largest value.

Finally, we remark on the second order $\epsilon$ expansion of the EMD near-horizon parametrization. Since the $\epsilon$ dependence is encapsulated in the $\mathfrak{c}$ coefficient in Eq. \eqref{r_horizonEMD}, it is important to study how this relation changes at second order. We have thus checked that for all BH models of interest, Eq.~\eqref{r_horizonEMD} does not receive corrections at $\order{\mathfrak{c}^2}$. Therefore, $\hr$ acquires $\order{\epsilon^2}$ contributions only from the expansion of $\mathfrak{c}=-\epsilon/(1+\epsilon)$. Moreover, the second-order expansion involves significantly more terms than the first-order one, namely linear and quadratic terms in $\{\theta_n\}$, as well as possible mixing among the linear coefficients. We find that, for a given order $(N,M)$ of the Padé approximation, the number of coefficients at second order scales quadratically with $K=N+M$ as $(K^2+5K+6)/2$. Similarly, at order $\epsilon^N$, the number of coefficients will grow as $\sim K^N$.

\newpage

\twocolumngrid

\bibliographystyle{unsrtnat}
\bibliography{refs.bib}

\end{document}